\documentclass[a4paper,11pt]{article}
\pdfoutput=1
\usepackage{jheppub}
\usepackage{graphicx}
\usepackage{bm}
\usepackage{slashed}
\usepackage{color}

    \title{Probing stop pair production at the LHC with graph neural networks}

    \author[a,b,c]{Murat Abdughani}
    \author[b,c]{Jie Ren}
    \author[a]{Lei Wu}
    \author[b,c,d]{Jin Min Yang}

    \affiliation[a]{Department of Physics and Institute of Theoretical Physics, Nanjing Normal University, Nanjing, 210023, China}
    \affiliation[b]{CAS Key Laboratory of Theoretical Physics, Institute of Theoretical Physics, Chinese Academy of Sciences, Beijing 100190, China}
    \affiliation[c]{School of Physics, University of Chinese Academy of Sciences, Beijing 100049, China}
    \affiliation[d]{Department of Physics, Tohoku University, Sendai 980-8578, Japan}

    \emailAdd{mulati@itp.ac.cn}
    \emailAdd{renjie@itp.ac.cn}
    \emailAdd{leiwu@itp.ac.cn}
    \emailAdd{jmyang@itp.ac.cn}

    \abstract{Top-squarks (stops) play a crucial role for the naturalness of supersymmetry (SUSY). However, searching for the stops is a tough task at the LHC. To dig the stops out of the huge LHC data, various expert-constructed kinematic variables or cutting-edge analysis techniques have been invented. In this paper, we propose to represent collision events as event graphs and use the message passing neutral network (MPNN) to analyze the events. As a proof-of-concept, we use our method in the search of the stop pair production at the LHC, and find that our MPNN can efficiently discriminate the signal and background events. In comparison with other machine learning methods (e.g. DNN), MPNN can enhance the mass reach of stop mass by several tens of GeV to over a hundred GeV.}

\begin{document}

    \maketitle


    \section{Introduction}
    After the discovery of the Higgs boson, the pursuit of new physics beyond the Standard Model (SM) is a primary goal of the LHC experiment. A major guideline in this endeavor is the naturalness principle which implies that the new physics for stabilizing the Higgs mass should appear at TeV scale. Among all the proposed scenarios, the weak scale SUSY remains as one of the most popular models, in which the quadratically divergent contribution to the Higgs mass from the top quark is canceled by the top-squarks (stops). Thus, the search for the stops is crucial for testing the naturalness of SUSY.

    However, searching for the stops at the LHC is a challenging task due to the complicated nature of super-particles (sparticles). For examples,
    (i) when $ m_{\tilde{t}_{1}} \gg  m_t + m_{\tilde{\chi}^0_1}$, the stop can decay to $t \tilde{\chi}^0_1$ and produce an energetic top quark. Using endpoint observables, like $M_T$ or $M_{T_2}$, the $t\bar{t}$ background can be efficiently reduced~\cite{Lester:1999tx, Barr:2003rg, Bai:2012gs, Cao:2012rz, Kilic:2012kw}.
    (ii) In the compressed region $ m_{\tilde{t}_1} - m_{\tilde{\chi}^0_1} \approx m_t$, the kinematics of stop pair events closely resemble the $t\bar{t}$ background events, rendering the searches rather difficult. Thanks to the ISR jet, the stop events in such a compressed region will have a peak-like feature around the ratio of missing transverse momentum vector to the transverse momentum vector of $t\bar{t}$ system~\cite{Hagiwara:2013tva, An:2015uwa, Macaluso:2015wja}, while the $t\bar{t}$ background does not show such a peak. If the LSP in the above compressed region becomes almost massless, the precision measurements of $t\bar{t}$ cross section~\cite{Czakon:2014fka} or spin-correlation~\cite{Han:2012fw} can also be used to probe the light stop.
    (iii) When the two body decays $\tilde{t}_1 \to t \tilde{\chi}^0_1$ and $\tilde{t}_1 \to b \tilde{\chi}^+_1$ are kinematically forbidden, the three-body decay $\tilde{t}_1\to W^+ b \tilde{\chi}^0_1$~\cite{Djouadi:2000bx}, the two-body flavor-changing decay $\tilde{t}_1 \to c \tilde{\chi}^0_1$~\cite{Han:2003qe, Muhlleitner:2011ww, Aebischer:2014lfa} or even the four-body decay $\tilde{t}_1 \to bf'\bar{f}\tilde{\chi}^0_1$~\cite{Boehm:1999tr} would happen. But due to the small mass splitting, the decay products of the stop are usually too soft to be observed. Thus the ISR/FSR jet (plus the heavy quark tagging) is needed to trigger these stop events~\cite{Ajaib:2011hs, Drees:2012dd, Yu:2012kj}.
    In addition, other miscellaneous cut-flow based studies of stop searches in different parameter space have been performed at the LHC~\cite{Perelstein:2008zt, Plehn:2012pr, Han:2013kga, Buckley:2014fqa, Goncalves:2014axa, Fuks:2014lva, Eifert:2014kea, Kobakhidze:2015dra, Hikasa:2015lma, Kobakhidze:2015scd, Cheng:2016mcw, Han:2016xet, Duan:2016vpp, Jackson:2016mfb, Goncalves:2016nil, Butter:2017cot, Kang:2017rfw, Duan:2017zar, Baer:2017pba}. A vast number of experimental searches for stop pair productions have been also devoted at the LHC in the past few years~\cite{Aaboud:2018zjf,Aaboud:2017nfd,Aaboud:2017ayj,Sirunyan:2017kqq,Sirunyan:2017xse,Sirunyan:2018omt}.
    Besides the traditional cut-flow based analysis technique, the machine learning (ML) method provides an alternative way for signal/background discrimination and new particle searches, which have been used for nearly three decades~\cite{Bhat:2010zz}. A great successful example in this direction is the use of Boosted Decision Trees~\cite{Roe:2004na} in the LHC experiment that led to the Higgs discovery. Recently, the ML techniques have been developed and applied for the studies of BSM phenomenology~\cite{Baldi:2014kfa, Baldi:2014pta, Bridges:2010de, Buckley:2011kc, Bornhauser:2013aya, Caron:2016hib, Bertone:2016mdy} increasingly.

    When using ML to deal with collision events, one has to first construct a representation of event and then choose a ML model to analyze that representation. An event is usually described by a set of particles with certain kinematic features. The geometrical relationship between these particles in the event is a sensitive probe to distinguish the signal and background events. In mathematics, such a geometrical pattern of a number of entities can be represented by a graph, which can be numerically analyzed by ML algorithms. Among them, the Message Passing Neural Networks (MPNNs)~\cite{2017arXiv170401212G} provide a general framework for supervised learning on graphs and are particularly suited for the problems where geometrical representation of patterns is to be learned. The MPNNs inherit the generality and powerfulness of the original Graph Neural Network (GNN)~\cite{Gori05, Scarselli09} and improve the training efficiency. They are non-linear models with a bunch of parameters that relates the output to the input graphs, where the supervised learning finds optimized parameters. The MPNNs have been applied to jet physics~\cite{2017arXiv170401212G} and other fields~\cite{Henrion17}.

    In this work, we first apply MPNN for the classification of signal and background events. Each HEP event is represented as an event graph, in which the nodes describe the final state particles and the edges reflect the geometric relations between each two nodes. In contrast with Deep Neural Network (DNN), MPNN is a dynamic neural network and intrinsically does not depend on the number and ordering of final state particles. Therefore, the MPNN is especially fit for processing the graph representation of collision event. As a proof-of-concept, we implement our method in the search of the stops through the process $pp \to \tilde{t}_1\tilde{t}^*_1 \to t\bar{t}\tilde{\chi}^0_1 \tilde{\chi}^0_1$ at the LHC.


    \section{Methodology}
        \begin{figure}[th]
    \center
        \includegraphics[width=14cm,height=7cm]{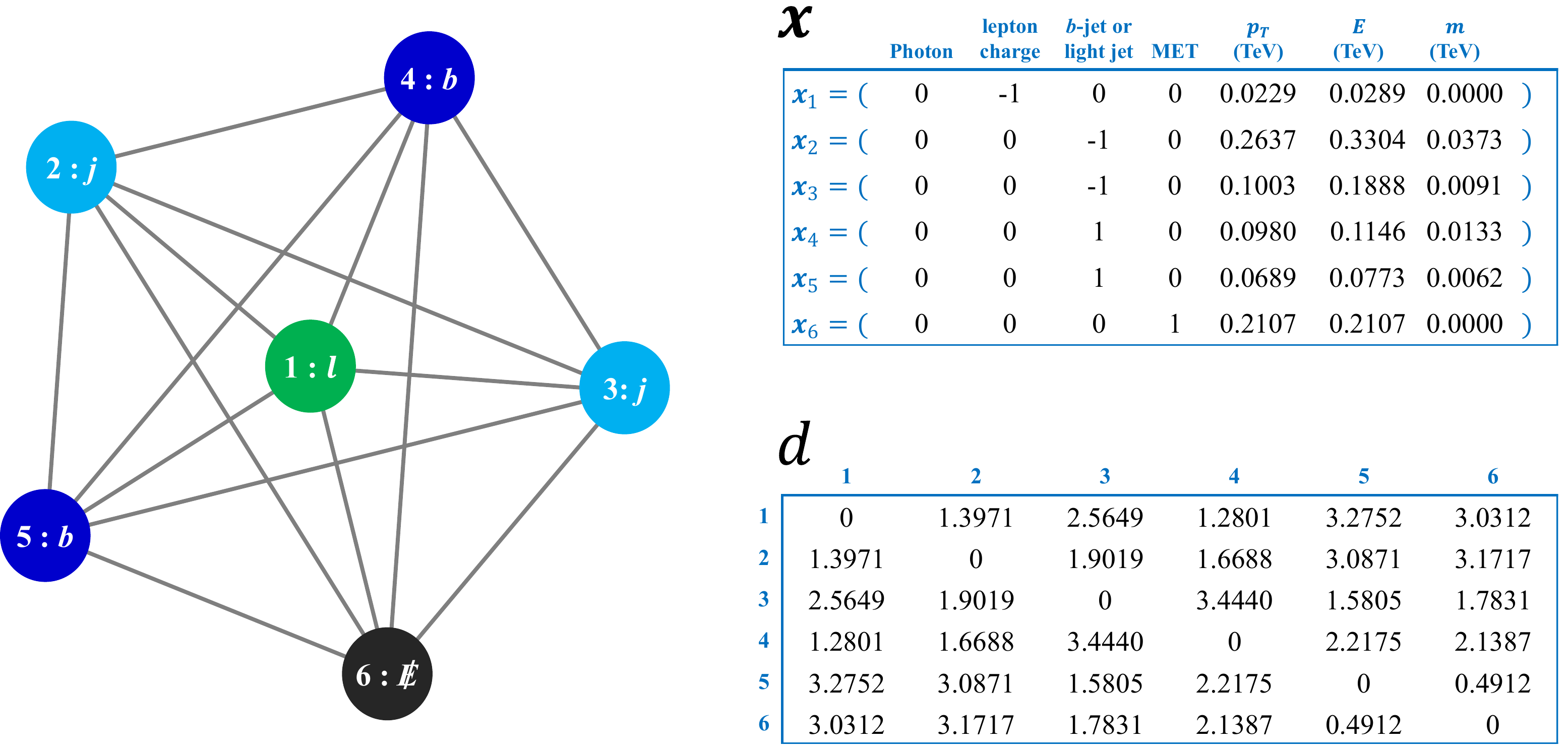}
        \caption{ An event graph with detailed node features and distance matrix, built from a Monte Carlo simulated event of the process $pp \to \tilde{t}_1 \tilde{t}^*_1 \to t\bar{t}\tilde{\chi}^0_1\tilde{\chi}^0_1 \to \ell + 2b + 2j + \slashed{E}_T$.}
        \label{event-graph}
    \end{figure}
    A collision event usually produces a number of final state objects which are reconstructed as photons, leptons and jets with four-momentum information. To construct an event graph, we represent each final state object as a node and connect each pair of nodes by an edge, which constitutes an undirected weighted complete graph. Each node $i$ has its feature vector $\bm{x}_i$ and each edge between node $i$ and $j$ has its weight matrix $d_{ij}$. As an illustration, we show an event graph of a Monte Carlo simulated event of the process $pp \to \tilde{t}_1 \tilde{t}^*_1 \to t\bar{t}\tilde{\chi}^0_1\tilde{\chi}^0_1 \to \ell + 2b + 2j + \slashed{E}_T$ in Fig.~\ref{event-graph}, where the detailed node feature vector $\bm{x}$ and edge weights matrix $d$ are also given. To be specific, the properties of the $i$-th final state are partly encoded into the 7-dimensional node feature vector $\bm{x}_i$. The first feature indicates that the final state is a photon (1) or not (0). If the final state is a lepton, its charge acts as the second feature; otherwise 0. The third feature indicates that the final state is a b-jet (1), light jet (-1) or not a jet (0). The fourth feature is used as an indicator of missing energy (MET) (1) or not (0). The rest three features are transverse momentum ($p_T$), energy ($E$) and mass ($m$) of the object. By construction, such a feature vector is compact. The edges are weighted by pair-distances $d_{ij} = \sqrt{\Delta y_{ij}^2 + \Delta\phi_{ij}^2}$ between $i$-th and $j$-th node in the graph, where $\Delta y$ and $\Delta \phi$ are the rapidity difference and azimuthal angle difference, respectively. It should be mentioned that the azimuthal angle $\phi$ is not encoded in the node features so that our graph representation is invariant under rotation in $\phi$, which is helpful for the stability of event classification.

    \begin{figure}[th]
    \center
        \includegraphics[width=14cm,height=7cm]{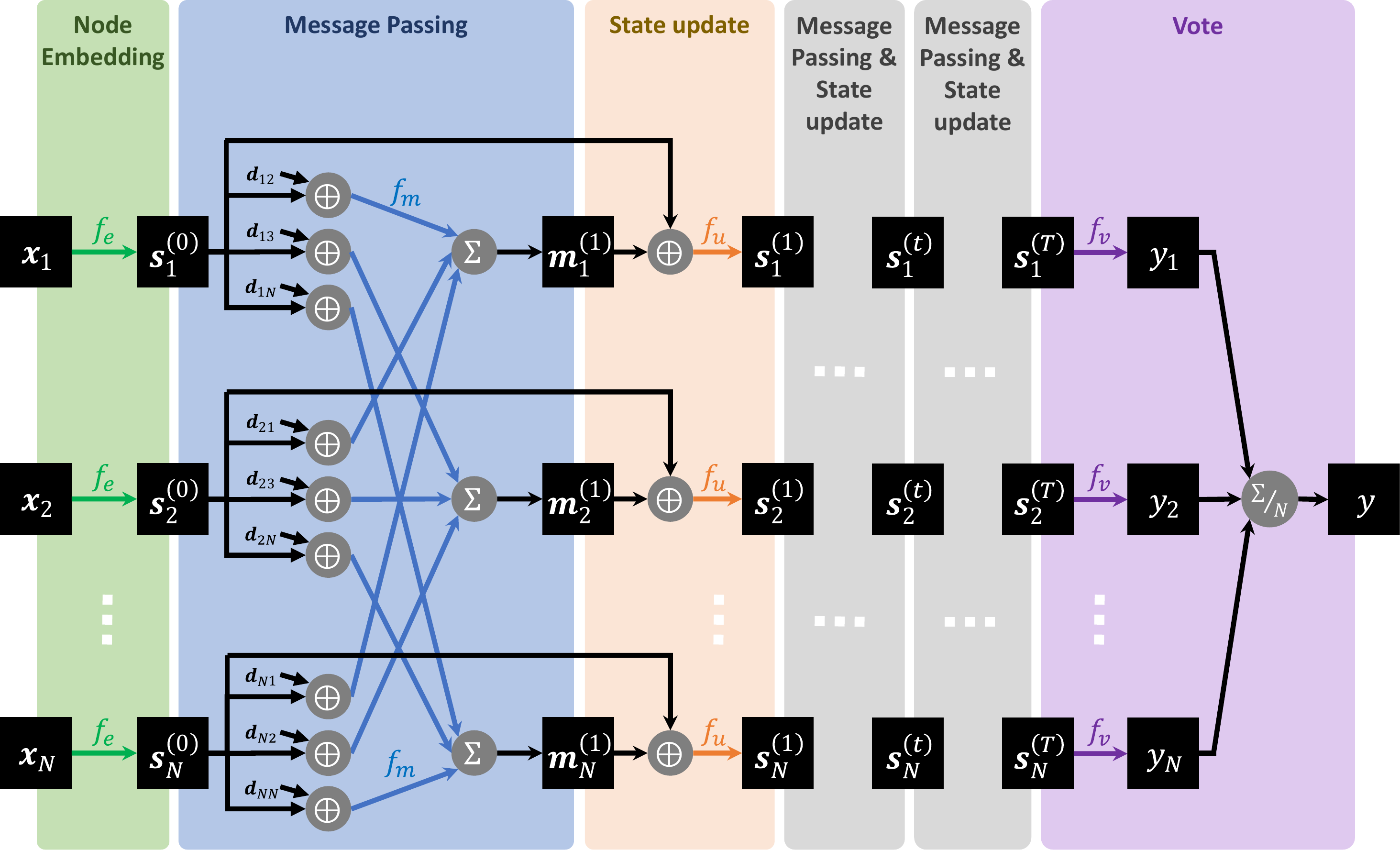}
        \caption{The architecture of our message passing neural network (MPNN) designed for event graph classification. The network is a stack of functional layers shown as shadowed blocks, which has $T$ pair of message passing and state update layers for automatic event feature extraction. State vectors $\bm{s}$, message vectors $\bm{m}$, votes $y_i$ and discrimination score $y$ are shown as black boxes. Colored arrows denote applying node embedding function $f_e$, message passing function $f_m$, state update function $f_u$ and vote function $f_v$, respectively. The operators are given in gray circles, where $\oplus$ performs vector concatenation, $\Sigma$ and $\Sigma/N$ are sum and average, respectively.}
        \label{mpnn}
    \end{figure}
    In this paper, we design a variant of MPNN to carry out the graph classification, whose architecture is presented in Fig.~\ref{mpnn}. Firstly, we embed each node features $\bm{x}_i$ into a higher dimensional state vector $\bm{s}_i^{(0)}$:
    \begin{equation}
    	\bm{s}_i^{(0)} = f_e(\bm{x}_i),
    \end{equation}
    where $f_e$ is called the node embedding function. The state vector $\bm{s}_i^{(0)}$ only encodes the $i$-th node features $\bm{x}_i$ without any information about the geometrical pattern of the graph. Then, the message passing techniques are utilized to perform event graph embedding, which will encode the whole event graph into each node state vector. At iteration $t$, each node $i$ collects the messages sent from other nodes $j$:
    \begin{equation}
    	\bm{m}_i^{(t)} = \sum_{j \neq i} \bm{m}_{i \leftarrow j}^{(t)} = \sum_{j \neq i} f_m^{(t)} (\bm{s}_{j}^{(t-1)}, d_{ij}),
    \end{equation}
    and update its state vector
    \begin{equation}
    	\bm{s}_i^{(t)} = f_u^{(t)} (\bm{s}_i^{(t-1)}, \bm{m}_i^{(t)}) ,
    \end{equation}
    where $f_m^{(t)}$ are the message functions and $f_u^{(t)}$ are the update functions. By repeating this procedure, the information in node states and the distances between nodes are disseminated with the sent messages, and each node updates its knowledge of other nodes and the relationships between all nodes. Therefore, after $T$ iterations, each resulting node state is an encoding of the whole graph, which is a compact representation of the information of both the kinematic features of all final states and the geometrical relationship between them. They are the event features that are automatically extracted from the input event graph. Next, each node votes a number as the likeness of the event to be signal-like, based on its own state vector,
    \begin{equation}
    	y_i = f_v(\bm{s}_i^{(T)}) ,
    \end{equation}
    where $f_v$ is the vote function. Finally, to make the prediction stable, we average the votes from each node
    \begin{equation}
    	y = \frac{1}{|\mathcal{V}|} \sum_i y_i
    \end{equation}
    as the final discrimination score of the event, where $|\mathcal{V}|$ is the number of nodes. The above operations form an end-to-end ML model, which maps event graphs directly to discrimination scores. The event selection can then be carried out by applying a specific cut $\theta_y$ on the score $y$; only events with $y > \theta_y$ will be selected out.

    In our following calculations, we use 30-dimensional state and message vectors, and choose single layer perceptrons as the node embedding, message passing, update and vote functions,
    \begin{eqnarray}
    	f_e(\bm{x}) &=& \mathrm{relu} \left( W_e \bm{x} + \bm{b}_e \right) , \\
    	f_m^{(t)}(\bm{s}, d) &=& \mathrm{relu} \left( W_m^{(t)} (\bm{s} \oplus [d]) + \bm{b}_m^{(t)} \right) , \\
    	f_u^{(t)}(\bm{s}, \bm{m}) &=& \mathrm{relu} \left( W_u^{(t)} (\bm{s} \oplus \bm{m}) + \bm{b}_u^{(t)} \right) , \\
    	f_v(\bm{s}) &=& \sigma \left( W_v \bm{s} + \bm{b}_v \right) ,
    \end{eqnarray}
    where $\oplus$ denotes vector concatenation, $\mathrm{relu}$ is the rectified linear unit, $\sigma$ is the sigmoid function, $W$s and $\bm{b}$s are learnable parameters. Independent message and update functions are used for each iteration $t$ up to $T=2$. Note that the pair distance $d_{ij}$ is only a real number. To ease the learning of the message functions, we represent the distance $d_{ij}$ in high dimensional vector space. Inspired from that radio basis function (RBF) networks can solve non-linear problems by mapping input into high dimensions using RBFs, we therefore choose the unnormalized Gaussian basis $\exp\{ (d - \mu_i)^2 / 2 \delta^2 \}$ as the RBFs to map the number $d_{ij}$ into a vector $[d]$. By design, the mean values $\mu_i$ of the RBFs are linearly distributed in the range of [0, 5] as $\mu_1 = 0, \mu_2 = 0.25, \mu_3 = 0.5, \cdots, \mu_{20} = 4.75, \mu_{21} = 5$, and RBFs have the same width of $\delta=0.25$. With such a method, we find that our result will become better, for example, the significance of benchmark point B can increase from 4.5 to 5.4 when $d_{ij}$ is expanded from 0 to 21. Based on our practice, above choices are a good trade-off between model complexity and prediction accuracy. Our MPNN model totally have 7051 learnable parameters.

    The MPNN can be efficiently trained using supervised learning techniques. We adopt binary-cross-entropy as the loss function. The Adam~\cite{DBLP:journals/corr/KingmaB14} optimizer with a learning rate of 0.001 is used to optimize the model parameters based on the gradients calculated on mini-batch of 500 training examples. A separate set of validation examples is used to measure the generalization performance while training to prevent over-fitting using the early-stopping technique. All these are implemented with the open-source deep learning framework PyTorch~\cite{pytorch} with strong GPU acceleration.

	Next, we compare the performance of MPNN with DNN in stop search at the LHC. Since the input size of DNN is fixed, we sort the nodes in each event graph by their identities and $p_T$ and then arrange the node features and edge weights into a fixed-size feature vector of the form
	\begin{equation}
		[ \bm{x}_{\slashed{E}_T}^T, \bm{x}_{\ell}^T, \bm{x}_{b_1}^T, \bm{x}_{b_2}^T, \bm{x}_{j_1}^T, \cdots, \bm{x}_{j_N}^T; d_{\slashed{E}_T \slashed{E}_T}, d_{\slashed{E}_T \ell}, \cdots, d_{j_N j_N} ],
		\label{input_for_dnn}
	\end{equation}
	where $N=$ 17 is the maximum number of light jets in our event graphs. Zero-padding is adopted to fill the missing values in the feature vector, namely for an
	event graph with $n$ light jets, $\bm{x}_{j_k} = \bm{0}\ (k > n)$ and $d_{j_k j_l} = 0\ (k\ \mathrm{or}\ l > n)$. 


    \section{Numerical results and discussions}
    As a proof-of-concept, we apply MPNN to investigate the observability of the stop through the process $pp \to \tilde{t}_1 \tilde{t}^*_1 \to t\bar{t}\tilde{\chi}^0_1\tilde{\chi}^0_1 \to \ell + 2b + 2j + \slashed E_T$ at 13 TeV LHC with the luminosity $\mathcal{L} = 36.1~\mathrm{fb}^{-1}$. We assume the LSP $\tilde{\chi}^0_1$ is pure bino and focus on the kinematic region of $m_{\tilde{t}_1} \ge m_t + m_{\tilde{\chi}^0_1}$. The dominant background events in this analysis arise from $t\bar{t}$, $W+\mathrm{jets}$ and $tW$. The $t\bar{t}Z(\to \nu \bar{\nu})$ background is non-negligible for a heavy stop and is included in our calculations as well. The multi-jet background can be estimated from data using a fake-factor method, which is found to be negligible in all regions\cite{Aaboud:2017aeu}.

    We use the event generator \textsf{MadGraph5\_aMC@NLO}~\cite{Alwall:2014hca} to simulate the signal and background events at the parton-level. Then we carry out the parton shower and hadronization with the \textsf{Pythia8.2}~\cite{Sjostrand:2014zea}. \textsf{Delphes-3.4.1}~\cite{deFavereau:2013fsa} is used for fast detector simulation. The anti-$k_t$ algorithm \cite{Cacciari:2008gp} with the distance parameter $R=0.4$ is chosen to cluster jets, and the $b$-tagging efficiency is assumed as 80\%. In the end, the event preselections are performed by \textsf{CheckMATE-2.0.14}~\cite{Drees:2013wra} using the following pre-selection cuts.
    \begin{itemize}
      \item We require exactly one lepton with $p_T(\ell) > 10$ GeV and $|\eta(\ell)| < 2.5$, and at least four jets with $p_T(j) > 25$ GeV and $|\eta| < 2.5$.
      \item We also require exactly two $b$-jets in the events.
      \item The transverse missing energy should satisfy $\slashed E_T >150$ GeV.
    \end{itemize}
      The NLO QCD corrected cross section of stop pair production is calculated with the \textsf{Prospino}~\cite{Beenakker:1999xh}. The $t\bar{t}$ and $W$+jets events are further normalized with their NNLO cross-sections, respectively~\cite{Czakon:2011xx, Boughezal:2015dva}. We evaluate the statistical significance with the formula,
       \begin{eqnarray}
         Z &=& S / \sqrt{B}.
       \end{eqnarray}
       To guarantee the statistics, we require at least 10 events after imposing cuts.

    \begin{table}[ht]
        \caption{The comparison of MPNN with DNN for significance ($Z$) of two benchmark points at 13 TeV LHC with the luminosity of ${\cal L}=36.1$ fb$^{-1}$. MPNN6 and DNN6 denote the results that use six objects (one lepton, two b-jets, two leading light-jets and MET) as inputs, respectively.}
        \label{comparison}
        \center
        \begin{tabular}{lp{5mm}rp{5mm}r}
            \hline\hline
            Benchmark && A && B \\
            \hline
            $m_{\tilde{t}_1}$ (GeV) && 525 && 900 \\
            $m_{\tilde{\chi}^0_1}$ (GeV) && 352 && 330 \\
            \hline
            $Z$(MPNN) && 4.6$\sigma$ && 5.4$\sigma$ \\
            $Z$(DNN) && 3.0$\sigma$ && 4.4$\sigma$ \\
            \hline
            $Z$(MPNN6) && 3.5$\sigma$ && 4.3$\sigma$ \\
			$Z$(DNN6) && 3.2$\sigma$ && 4.3$\sigma$ \\
            \hline\hline
        \end{tabular}
    \end{table}
In Table~\ref{comparison}, we choose two Benchmark Points(BPs) with distinctive kinematic features to compare the performance of MPNN and DNN. The BP-A lies in the compressed region with $m_{\tilde{t}_1} \approx m_t + m_{\tilde{\chi}^0_1}$, while the BP-B lies in the uncompressed region with $m_{\tilde{t}_1} \gg m_t + m_{\tilde{\chi}^0_1}$. For BP-$A/B$, we generated 14/4 million signal events and 100 million background events. After the pre-selection, 300,000 signal events and 300,000 background events are collected as training examples. We also collect a separate set of 100,000 signal events and 100,000 background events as validation examples. In addition, we also show the results of MPNN6 and DNN6, which use six objects (one lepton, two b-jets, two leading light-jets and MET) used as inputs, respectively. The reduced event graphs for MPNN6 have 6 nodes and 36 edges. The input feature vectors for DNN6 are of the form of Eq.~(\ref{input_for_dnn}) with $N=2$ so that it has 78 input features.

In order to optimize the DNN results, we perform a grid scan of hyperparameters for BP-A and B. 
In our scan, we find that the significance generally increases with more hidden layers and neurons per hidden layer. But the increasing complexity of the DNN model limits the improvement. The best DNN model for BP-A/B has 7/4 hidden layers, 700/300 neurons per hidden layer and a dropout ratio of 0.5/0.4, which owns 3357201/447901 learning parameters and gives the best significance of 3.0/4.4$\sigma$. The same optimization procedure is adopted for DNN6 model in Table~\ref{comparison} and also for each sample in Fig.~\ref{exclusion}.

From Table~\ref{comparison}, we can see that the MPNN method have better significance than DNN for both BPs, in particular for the BP-A with $m_{\tilde{t}_1}\approx m_{\tilde{\chi}^0_1}+m_{t}$. The significance of BP-A increases from $3.2\sigma$ (DNN6) to $3.5\sigma$ (MPNN6), and $3.0\sigma$ (DNN) to $4.6\sigma$ (MPNN). With more input features, MPNN outperforms over MPNN6, while DNN has no such a feature, since more learnable parameters usually leads to more serious over-fitting for DNN.

    \begin{figure}[ht]
    \center
        \includegraphics[width=14cm,height=14cm]{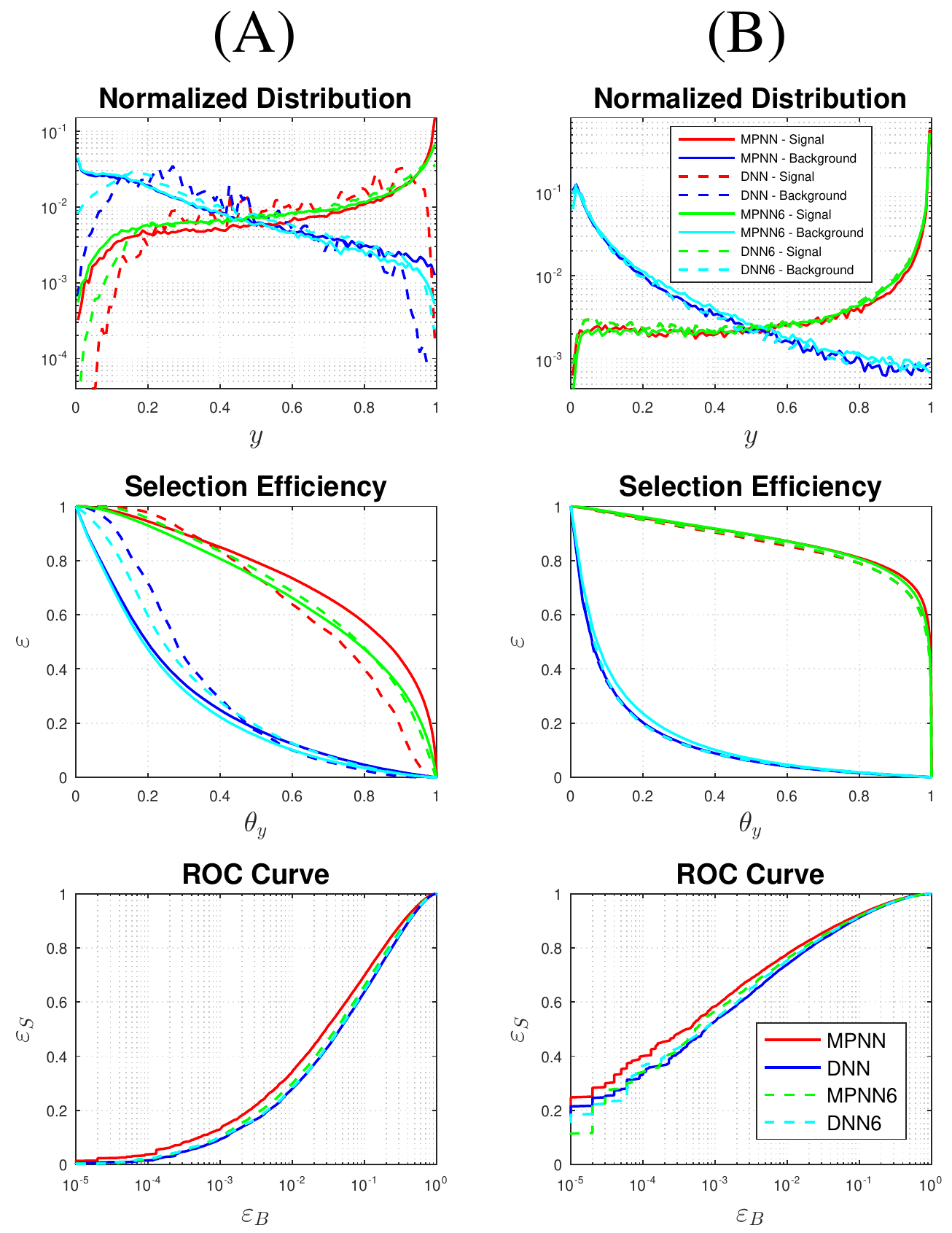}
        \caption{The discriminating power of MPNN(6) and DNN(6) on signal and background for benchmark points A (left panel) and B (right panel) in Table~\ref{comparison}. The top panel is the distributions of discrimination score $y$ for signal and background events. The middle panel is the selection efficiency $\epsilon$ versus the cut threshold $\theta_y$. The bottom panel shows ROC curves of the MPNN(6) and DNN(6).}
        \label{benchmark}
    \end{figure}
    In Fig.~\ref{benchmark}, we further present the discriminating power of MPNNs and DNNs on signal and background events for benchmark points A and B. Since the over-fitting is very small, we only show the results on validation set. From the top panel, we can see that the signal and background events are well separated in the distributions of the discrimination score for both of MPNNs and DNNs. But for MPNNs, the score of signals are inclined to have larger value than DNNs, while the score of backgrounds have smaller values than DNNs. From the middle panel, we can find that MPNNs has higher signal selection efficiency $\varepsilon_S$ and lower background selection efficiency $\varepsilon_B$ than DNNs. This leads to MPNNs having sharper receiver operating characteristic (ROC) curves than DNNs, as shown in the bottom panel of Fig.~\ref{benchmark}.

    \begin{figure}[ht]
    \center
        \includegraphics[width=14cm,height=10cm]{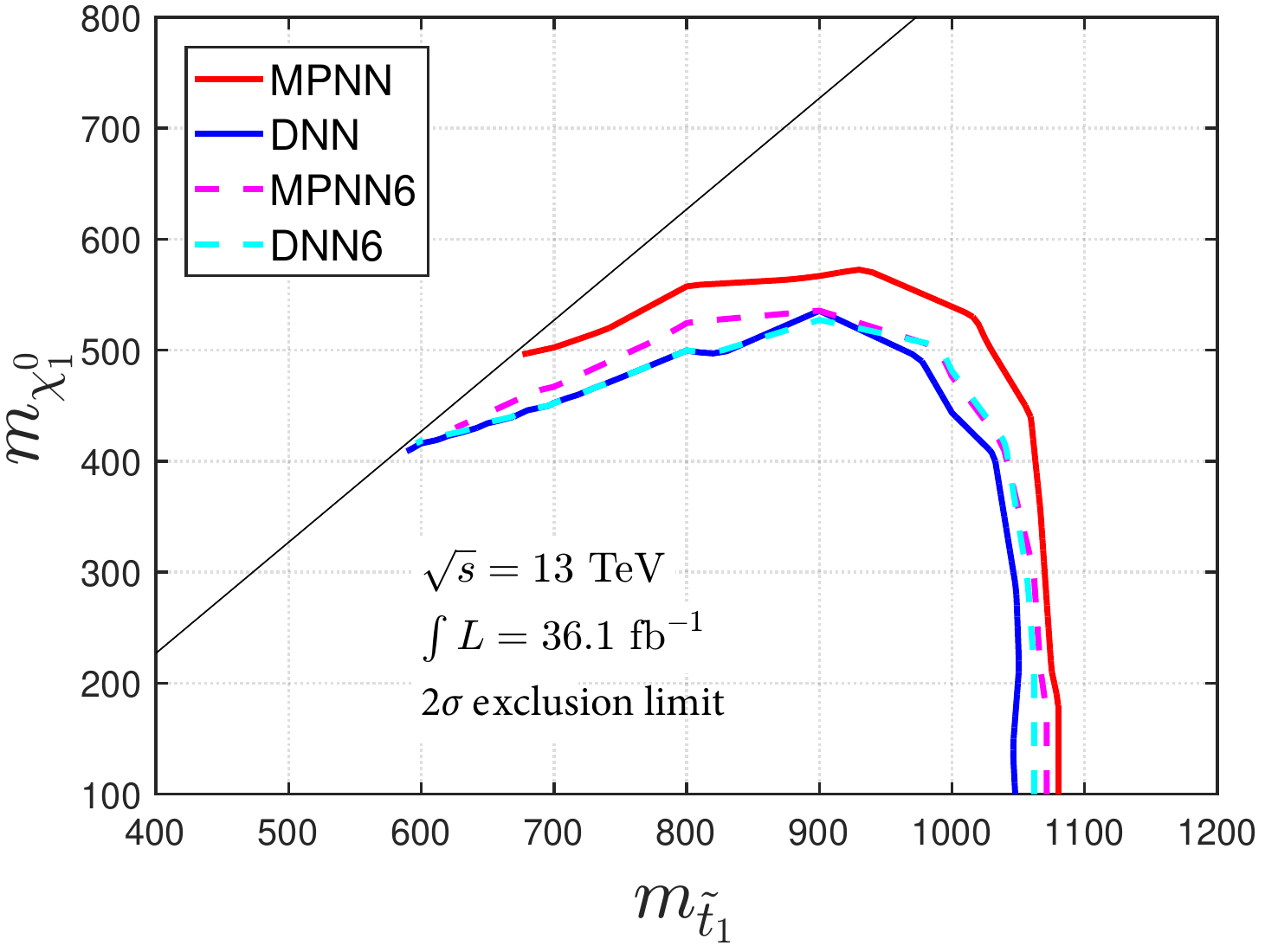}
        \caption{Exclusion limits on the plane of $m_{\tilde{t}_1}$ versus $m_{\tilde{\chi}^0_1}$ by using MPNN(6) and DNN(6) at 13 TeV LHC with the luminosity of ${\cal L}=36.1$ fb$^{-1}$.}
        \label{exclusion}
    \end{figure}
    In Fig.~5, we perform a grid scan on the plane of $m_{\tilde{t}_1}$ versus $m_{\tilde{\chi}^0_1}$ by using MPNN(6) and DNN(6) at 13 TeV LHC with the luminosity ${\cal L}=36.1$ fb$^{-1}$. We take the right-handed $\tilde{t}_1$ as the example and simulate 40 parameter points in our grid scan. The pre-selections for signal and background events are the same as those for the benchmark points in Tab.~1. We optimize the DNN(6) and MPNN(6) models for each parameter point as the approaches used for benchmarks points. As a theoretical work, we define the exclusion limit when the significance $Z=2\sigma$. Then, the exclusion line is drawn by using bilinear interpolation. We can see that the MPNN method can produce a stronger exclusion limit than the DNN method. For examples, in the compressed region with $m_{\tilde{t}_1} \approx m_{\tilde{\chi}^0_1} + m_{t}$, the stop mass reach from MPNN can be 670 GeV, which is about 90 GeV larger than that from the DNN. In other regions with $m_{\tilde{t}_1} > m_{\tilde{\chi}^0_1} + m_{t}$, the exclusion limit on stop mass for given neutralino mass from the MPNN can be greater than DNN by about several tens of GeV to more than one hundred GeV. Unlike the MPNN, the results of the MPNN6 are not much better than that of the DNN6. This is because that the man-made truncation on the number of final states in the MPNN6 will badly worsen its performance. In the MPNN6, the kinematical information coded in the event graphs for some parameter points having large multiplicity are incomplete. Thus the full correlations among final states, which is the most important factor for discriminating signal and backgrounds in the MPNN method, are broken. On the other hand, the performance of DNN seems worse than that of DNN6 in most of region in Fig.~\ref{exclusion}. This is because the DNN involves much more training parameters so that it has poorer generalization capability. Therefore, we expect that the MPNN approach will have better performance than the DNN approach when the physical processes have high-multiplicity final states at the LHC.


    \section{Conclusions}
    In this paper, we proposed to represent a collision event in high energy physics as an event graph with a set of nodes and edges, and use the Message Passing Neural Networks to deal with the problems of discrimination of signal and background events at colliders. As a proof of concept, we applied our approach to the search for the stop pair production at the LHC. We found that the geometrical pattern of the event that depends on correlations among final states can be a sensitive feature in the MPNN to discriminate signal and background events. Furthermore, we compared the performance of our MPNN with the DNN, and found that our MPNN can enhance the mass reach of stop mass by several tens of GeV to over a hundred GeV.\\


    \textbf{\textit{Acknowledgment.}}
    Part of this work was done while M. A. was visiting Nanjing Normal University. This work was supported by the National Natural Science Foundation of China (NNSFC) under grant Nos. 11705093, 11675242 and 11851303, by Peng-Huan-Wu Theoretical Physics Innovation Center (11747601), by the CAS Center for Excellence in Particle Physics (CCEPP), by the CAS Key Research Program of Frontier Sciences and by a Key R\&D Program of Ministry of Science and Technology under number 2017YFA0402200-04.



\begin{thebibliography}{67}%
\makeatletter
\providecommand \@ifxundefined [1]{%
 \@ifx{#1\undefined}
}%
\providecommand \@ifnum [1]{%
 \ifnum #1\expandafter \@firstoftwo
 \else \expandafter \@secondoftwo
 \fi
}%
\providecommand \@ifx [1]{%
 \ifx #1\expandafter \@firstoftwo
 \else \expandafter \@secondoftwo
 \fi
}%
\providecommand \natexlab [1]{#1}%
\providecommand \enquote  [1]{``#1''}%
\providecommand \bibnamefont  [1]{#1}%
\providecommand \bibfnamefont [1]{#1}%
\providecommand \citenamefont [1]{#1}%
\providecommand \href@noop [0]{\@secondoftwo}%
\providecommand \href [0]{\begingroup \@sanitize@url \@href}%
\providecommand \@href[1]{\@@startlink{#1}\@@href}%
\providecommand \@@href[1]{\endgroup#1\@@endlink}%
\providecommand \@sanitize@url [0]{\catcode `\\12\catcode `\$12\catcode
  `\&12\catcode `\#12\catcode `\^12\catcode `\_12\catcode `\%12\relax}%
\providecommand \@@startlink[1]{}%
\providecommand \@@endlink[0]{}%
\providecommand \url  [0]{\begingroup\@sanitize@url \@url }%
\providecommand \@url [1]{\endgroup\@href {#1}{\urlprefix }}%
\providecommand \urlprefix  [0]{URL }%
\providecommand \Eprint [0]{\href }%
\providecommand \doibase [0]{http://dx.doi.org/}%
\providecommand \selectlanguage [0]{\@gobble}%
\providecommand \bibinfo  [0]{\@secondoftwo}%
\providecommand \bibfield  [0]{\@secondoftwo}%
\providecommand \translation [1]{[#1]}%
\providecommand \BibitemOpen [0]{}%
\providecommand \bibitemStop [0]{}%
\providecommand \bibitemNoStop [0]{.\EOS\space}%
\providecommand \EOS [0]{\spacefactor3000\relax}%
\providecommand \BibitemShut  [1]{\csname bibitem#1\endcsname}%
\let\auto@bib@innerbib\@empty
\bibitem [{\citenamefont {Lester}\ and\ \citenamefont
  {Summers}(1999)}]{Lester:1999tx}%
  \BibitemOpen
  \bibfield  {author} {\bibinfo {author} {\bibfnamefont {C.~G.}\ \bibnamefont
  {Lester}}\ and\ \bibinfo {author} {\bibfnamefont {D.~J.}\ \bibnamefont
  {Summers}},\ }\href {\doibase 10.1016/S0370-2693(99)00945-4} {\bibfield
  {journal} {\bibinfo  {journal} {Phys. Lett.}\ }\textbf {\bibinfo {volume}
  {B463}},\ \bibinfo {pages} {99} (\bibinfo {year} {1999})},\ \Eprint
  {http://arxiv.org/abs/hep-ph/9906349} {arXiv:hep-ph/9906349 [hep-ph]}
  \BibitemShut {NoStop}%
\bibitem [{\citenamefont {Barr}\ \emph {et~al.}(2003)\citenamefont {Barr},
  \citenamefont {Lester},\ and\ \citenamefont {Stephens}}]{Barr:2003rg}%
  \BibitemOpen
  \bibfield  {author} {\bibinfo {author} {\bibfnamefont {A.}~\bibnamefont
  {Barr}}, \bibinfo {author} {\bibfnamefont {C.}~\bibnamefont {Lester}}, \ and\
  \bibinfo {author} {\bibfnamefont {P.}~\bibnamefont {Stephens}},\ }\href
  {\doibase 10.1088/0954-3899/29/10/304} {\bibfield  {journal} {\bibinfo
  {journal} {J. Phys.}\ }\textbf {\bibinfo {volume} {G29}},\ \bibinfo {pages}
  {2343} (\bibinfo {year} {2003})},\ \Eprint
  {http://arxiv.org/abs/hep-ph/0304226} {arXiv:hep-ph/0304226 [hep-ph]}
  \BibitemShut {NoStop}%
\bibitem [{\citenamefont {Bai}\ \emph {et~al.}(2012)\citenamefont {Bai},
  \citenamefont {Cheng}, \citenamefont {Gallicchio},\ and\ \citenamefont
  {Gu}}]{Bai:2012gs}%
  \BibitemOpen
  \bibfield  {author} {\bibinfo {author} {\bibfnamefont {Y.}~\bibnamefont
  {Bai}}, \bibinfo {author} {\bibfnamefont {H.-C.}\ \bibnamefont {Cheng}},
  \bibinfo {author} {\bibfnamefont {J.}~\bibnamefont {Gallicchio}}, \ and\
  \bibinfo {author} {\bibfnamefont {J.}~\bibnamefont {Gu}},\ }\href {\doibase
  10.1007/JHEP07(2012)110} {\bibfield  {journal} {\bibinfo  {journal} {JHEP}\
  }\textbf {\bibinfo {volume} {07}},\ \bibinfo {pages} {110} (\bibinfo {year}
  {2012})},\ \Eprint {http://arxiv.org/abs/1203.4813} {arXiv:1203.4813
  [hep-ph]} \BibitemShut {NoStop}%
\bibitem [{\citenamefont {Cao}\ \emph {et~al.}(2012)\citenamefont {Cao},
  \citenamefont {Han}, \citenamefont {Wu}, \citenamefont {Yang},\ and\
  \citenamefont {Zhang}}]{Cao:2012rz}%
  \BibitemOpen
  \bibfield  {author} {\bibinfo {author} {\bibfnamefont {J.}~\bibnamefont
  {Cao}}, \bibinfo {author} {\bibfnamefont {C.}~\bibnamefont {Han}}, \bibinfo
  {author} {\bibfnamefont {L.}~\bibnamefont {Wu}}, \bibinfo {author}
  {\bibfnamefont {J.~M.}\ \bibnamefont {Yang}}, \ and\ \bibinfo {author}
  {\bibfnamefont {Y.}~\bibnamefont {Zhang}},\ }\href {\doibase
  10.1007/JHEP11(2012)039} {\bibfield  {journal} {\bibinfo  {journal} {JHEP}\
  }\textbf {\bibinfo {volume} {11}},\ \bibinfo {pages} {039} (\bibinfo {year}
  {2012})},\ \Eprint {http://arxiv.org/abs/1206.3865} {arXiv:1206.3865
  [hep-ph]} \BibitemShut {NoStop}%
\bibitem [{\citenamefont {Kilic}\ and\ \citenamefont
  {Tweedie}(2013)}]{Kilic:2012kw}%
  \BibitemOpen
  \bibfield  {author} {\bibinfo {author} {\bibfnamefont {C.}~\bibnamefont
  {Kilic}}\ and\ \bibinfo {author} {\bibfnamefont {B.}~\bibnamefont
  {Tweedie}},\ }\href {\doibase 10.1007/JHEP04(2013)110} {\bibfield  {journal}
  {\bibinfo  {journal} {JHEP}\ }\textbf {\bibinfo {volume} {04}},\ \bibinfo
  {pages} {110} (\bibinfo {year} {2013})},\ \Eprint
  {http://arxiv.org/abs/1211.6106} {arXiv:1211.6106 [hep-ph]} \BibitemShut
  {NoStop}%
\bibitem [{\citenamefont {Hagiwara}\ and\ \citenamefont
  {Yamada}(2015)}]{Hagiwara:2013tva}%
  \BibitemOpen
  \bibfield  {author} {\bibinfo {author} {\bibfnamefont {K.}~\bibnamefont
  {Hagiwara}}\ and\ \bibinfo {author} {\bibfnamefont {T.}~\bibnamefont
  {Yamada}},\ }\href {\doibase 10.1103/PhysRevD.91.094007} {\bibfield
  {journal} {\bibinfo  {journal} {Phys. Rev.}\ }\textbf {\bibinfo {volume}
  {D91}},\ \bibinfo {pages} {094007} (\bibinfo {year} {2015})},\ \Eprint
  {http://arxiv.org/abs/1307.1553} {arXiv:1307.1553 [hep-ph]} \BibitemShut
  {NoStop}%
\bibitem [{\citenamefont {An}\ and\ \citenamefont {Wang}(2015)}]{An:2015uwa}%
  \BibitemOpen
  \bibfield  {author} {\bibinfo {author} {\bibfnamefont {H.}~\bibnamefont
  {An}}\ and\ \bibinfo {author} {\bibfnamefont {L.-T.}\ \bibnamefont {Wang}},\
  }\href {\doibase 10.1103/PhysRevLett.115.181602} {\bibfield  {journal}
  {\bibinfo  {journal} {Phys. Rev. Lett.}\ }\textbf {\bibinfo {volume} {115}},\
  \bibinfo {pages} {181602} (\bibinfo {year} {2015})},\ \Eprint
  {http://arxiv.org/abs/1506.00653} {arXiv:1506.00653 [hep-ph]} \BibitemShut
  {NoStop}%
\bibitem [{\citenamefont {Macaluso}\ \emph {et~al.}(2016)\citenamefont
  {Macaluso}, \citenamefont {Park}, \citenamefont {Shih},\ and\ \citenamefont
  {Tweedie}}]{Macaluso:2015wja}%
  \BibitemOpen
  \bibfield  {author} {\bibinfo {author} {\bibfnamefont {S.}~\bibnamefont
  {Macaluso}}, \bibinfo {author} {\bibfnamefont {M.}~\bibnamefont {Park}},
  \bibinfo {author} {\bibfnamefont {D.}~\bibnamefont {Shih}}, \ and\ \bibinfo
  {author} {\bibfnamefont {B.}~\bibnamefont {Tweedie}},\ }\href {\doibase
  10.1007/JHEP03(2016)151} {\bibfield  {journal} {\bibinfo  {journal} {JHEP}\
  }\textbf {\bibinfo {volume} {03}},\ \bibinfo {pages} {151} (\bibinfo {year}
  {2016})},\ \Eprint {http://arxiv.org/abs/1506.07885} {arXiv:1506.07885
  [hep-ph]} \BibitemShut {NoStop}%
\bibitem [{\citenamefont {Czakon}\ \emph {et~al.}(2014)\citenamefont {Czakon},
  \citenamefont {Mitov}, \citenamefont {Papucci}, \citenamefont {Ruderman},\
  and\ \citenamefont {Weiler}}]{Czakon:2014fka}%
  \BibitemOpen
  \bibfield  {author} {\bibinfo {author} {\bibfnamefont {M.}~\bibnamefont
  {Czakon}}, \bibinfo {author} {\bibfnamefont {A.}~\bibnamefont {Mitov}},
  \bibinfo {author} {\bibfnamefont {M.}~\bibnamefont {Papucci}}, \bibinfo
  {author} {\bibfnamefont {J.~T.}\ \bibnamefont {Ruderman}}, \ and\ \bibinfo
  {author} {\bibfnamefont {A.}~\bibnamefont {Weiler}},\ }\href {\doibase
  10.1103/PhysRevLett.113.201803} {\bibfield  {journal} {\bibinfo  {journal}
  {Phys. Rev. Lett.}\ }\textbf {\bibinfo {volume} {113}},\ \bibinfo {pages}
  {201803} (\bibinfo {year} {2014})},\ \Eprint {http://arxiv.org/abs/1407.1043}
  {arXiv:1407.1043 [hep-ph]} \BibitemShut {NoStop}%
\bibitem [{\citenamefont {Han}\ \emph {et~al.}(2012)\citenamefont {Han},
  \citenamefont {Katz}, \citenamefont {Krohn},\ and\ \citenamefont
  {Reece}}]{Han:2012fw}%
  \BibitemOpen
  \bibfield  {author} {\bibinfo {author} {\bibfnamefont {Z.}~\bibnamefont
  {Han}}, \bibinfo {author} {\bibfnamefont {A.}~\bibnamefont {Katz}}, \bibinfo
  {author} {\bibfnamefont {D.}~\bibnamefont {Krohn}}, \ and\ \bibinfo {author}
  {\bibfnamefont {M.}~\bibnamefont {Reece}},\ }\href {\doibase
  10.1007/JHEP08(2012)083} {\bibfield  {journal} {\bibinfo  {journal} {JHEP}\
  }\textbf {\bibinfo {volume} {08}},\ \bibinfo {pages} {083} (\bibinfo {year}
  {2012})},\ \Eprint {http://arxiv.org/abs/1205.5808} {arXiv:1205.5808
  [hep-ph]} \BibitemShut {NoStop}%
\bibitem [{\citenamefont {Djouadi}\ and\ \citenamefont
  {Mambrini}(2001)}]{Djouadi:2000bx}%
  \BibitemOpen
  \bibfield  {author} {\bibinfo {author} {\bibfnamefont {A.}~\bibnamefont
  {Djouadi}}\ and\ \bibinfo {author} {\bibfnamefont {Y.}~\bibnamefont
  {Mambrini}},\ }\href {\doibase 10.1103/PhysRevD.63.115005} {\bibfield
  {journal} {\bibinfo  {journal} {Phys. Rev.}\ }\textbf {\bibinfo {volume}
  {D63}},\ \bibinfo {pages} {115005} (\bibinfo {year} {2001})},\ \Eprint
  {http://arxiv.org/abs/hep-ph/0011364} {arXiv:hep-ph/0011364 [hep-ph]}
  \BibitemShut {NoStop}%
\bibitem [{\citenamefont {Han}\ \emph {et~al.}(2004)\citenamefont {Han},
  \citenamefont {Hikasa}, \citenamefont {Yang},\ and\ \citenamefont
  {Zhang}}]{Han:2003qe}%
  \BibitemOpen
  \bibfield  {author} {\bibinfo {author} {\bibfnamefont {T.}~\bibnamefont
  {Han}}, \bibinfo {author} {\bibfnamefont {K.-i.}\ \bibnamefont {Hikasa}},
  \bibinfo {author} {\bibfnamefont {J.~M.}\ \bibnamefont {Yang}}, \ and\
  \bibinfo {author} {\bibfnamefont {X.-m.}\ \bibnamefont {Zhang}},\ }\href
  {\doibase 10.1103/PhysRevD.70.055001} {\bibfield  {journal} {\bibinfo
  {journal} {Phys. Rev.}\ }\textbf {\bibinfo {volume} {D70}},\ \bibinfo {pages}
  {055001} (\bibinfo {year} {2004})},\ \Eprint
  {http://arxiv.org/abs/hep-ph/0312129} {arXiv:hep-ph/0312129 [hep-ph]}
  \BibitemShut {NoStop}%
\bibitem [{\citenamefont {Muhlleitner}\ and\ \citenamefont
  {Popenda}(2011)}]{Muhlleitner:2011ww}%
  \BibitemOpen
  \bibfield  {author} {\bibinfo {author} {\bibfnamefont {M.}~\bibnamefont
  {Muhlleitner}}\ and\ \bibinfo {author} {\bibfnamefont {E.}~\bibnamefont
  {Popenda}},\ }\href {\doibase 10.1007/JHEP04(2011)095} {\bibfield  {journal}
  {\bibinfo  {journal} {JHEP}\ }\textbf {\bibinfo {volume} {04}},\ \bibinfo
  {pages} {095} (\bibinfo {year} {2011})},\ \Eprint
  {http://arxiv.org/abs/1102.5712} {arXiv:1102.5712 [hep-ph]} \BibitemShut
  {NoStop}%
\bibitem [{\citenamefont {Aebischer}\ \emph {et~al.}(2015)\citenamefont
  {Aebischer}, \citenamefont {Crivellin},\ and\ \citenamefont
  {Greub}}]{Aebischer:2014lfa}%
  \BibitemOpen
  \bibfield  {author} {\bibinfo {author} {\bibfnamefont {J.}~\bibnamefont
  {Aebischer}}, \bibinfo {author} {\bibfnamefont {A.}~\bibnamefont
  {Crivellin}}, \ and\ \bibinfo {author} {\bibfnamefont {C.}~\bibnamefont
  {Greub}},\ }\href {\doibase 10.1103/PhysRevD.91.035010} {\bibfield  {journal}
  {\bibinfo  {journal} {Phys. Rev.}\ }\textbf {\bibinfo {volume} {D91}},\
  \bibinfo {pages} {035010} (\bibinfo {year} {2015})},\ \Eprint
  {http://arxiv.org/abs/1410.8459} {arXiv:1410.8459 [hep-ph]} \BibitemShut
  {NoStop}%
\bibitem [{\citenamefont {Boehm}\ \emph {et~al.}(2000)\citenamefont {Boehm},
  \citenamefont {Djouadi},\ and\ \citenamefont {Mambrini}}]{Boehm:1999tr}%
  \BibitemOpen
  \bibfield  {author} {\bibinfo {author} {\bibfnamefont {C.}~\bibnamefont
  {Boehm}}, \bibinfo {author} {\bibfnamefont {A.}~\bibnamefont {Djouadi}}, \
  and\ \bibinfo {author} {\bibfnamefont {Y.}~\bibnamefont {Mambrini}},\ }\href
  {\doibase 10.1103/PhysRevD.61.095006} {\bibfield  {journal} {\bibinfo
  {journal} {Phys. Rev.}\ }\textbf {\bibinfo {volume} {D61}},\ \bibinfo {pages}
  {095006} (\bibinfo {year} {2000})},\ \Eprint
  {http://arxiv.org/abs/hep-ph/9907428} {arXiv:hep-ph/9907428 [hep-ph]}
  \BibitemShut {NoStop}%
\bibitem [{\citenamefont {Ajaib}\ \emph {et~al.}(2012)\citenamefont {Ajaib},
  \citenamefont {Li},\ and\ \citenamefont {Shafi}}]{Ajaib:2011hs}%
  \BibitemOpen
  \bibfield  {author} {\bibinfo {author} {\bibfnamefont {M.~A.}\ \bibnamefont
  {Ajaib}}, \bibinfo {author} {\bibfnamefont {T.}~\bibnamefont {Li}}, \ and\
  \bibinfo {author} {\bibfnamefont {Q.}~\bibnamefont {Shafi}},\ }\href
  {\doibase 10.1103/PhysRevD.85.055021} {\bibfield  {journal} {\bibinfo
  {journal} {Phys. Rev.}\ }\textbf {\bibinfo {volume} {D85}},\ \bibinfo {pages}
  {055021} (\bibinfo {year} {2012})},\ \Eprint {http://arxiv.org/abs/1111.4467}
  {arXiv:1111.4467 [hep-ph]} \BibitemShut {NoStop}%
\bibitem [{\citenamefont {Drees}\ \emph {et~al.}(2012)\citenamefont {Drees},
  \citenamefont {Hanussek},\ and\ \citenamefont {Kim}}]{Drees:2012dd}%
  \BibitemOpen
  \bibfield  {author} {\bibinfo {author} {\bibfnamefont {M.}~\bibnamefont
  {Drees}}, \bibinfo {author} {\bibfnamefont {M.}~\bibnamefont {Hanussek}}, \
  and\ \bibinfo {author} {\bibfnamefont {J.~S.}\ \bibnamefont {Kim}},\ }\href
  {\doibase 10.1103/PhysRevD.86.035024} {\bibfield  {journal} {\bibinfo
  {journal} {Phys. Rev.}\ }\textbf {\bibinfo {volume} {D86}},\ \bibinfo {pages}
  {035024} (\bibinfo {year} {2012})},\ \Eprint {http://arxiv.org/abs/1201.5714}
  {arXiv:1201.5714 [hep-ph]} \BibitemShut {NoStop}%
\bibitem [{\citenamefont {Yu}\ \emph {et~al.}(2013)\citenamefont {Yu},
  \citenamefont {Bi}, \citenamefont {Yan},\ and\ \citenamefont
  {Yin}}]{Yu:2012kj}%
  \BibitemOpen
  \bibfield  {author} {\bibinfo {author} {\bibfnamefont {Z.-H.}\ \bibnamefont
  {Yu}}, \bibinfo {author} {\bibfnamefont {X.-J.}\ \bibnamefont {Bi}}, \bibinfo
  {author} {\bibfnamefont {Q.-S.}\ \bibnamefont {Yan}}, \ and\ \bibinfo
  {author} {\bibfnamefont {P.-F.}\ \bibnamefont {Yin}},\ }\href {\doibase
  10.1103/PhysRevD.87.055007} {\bibfield  {journal} {\bibinfo  {journal} {Phys.
  Rev.}\ }\textbf {\bibinfo {volume} {D87}},\ \bibinfo {pages} {055007}
  (\bibinfo {year} {2013})},\ \Eprint {http://arxiv.org/abs/1211.2997}
  {arXiv:1211.2997 [hep-ph]} \BibitemShut {NoStop}%
\bibitem [{\citenamefont {Perelstein}\ and\ \citenamefont
  {Weiler}(2009)}]{Perelstein:2008zt}%
  \BibitemOpen
  \bibfield  {author} {\bibinfo {author} {\bibfnamefont {M.}~\bibnamefont
  {Perelstein}}\ and\ \bibinfo {author} {\bibfnamefont {A.}~\bibnamefont
  {Weiler}},\ }\href {\doibase 10.1088/1126-6708/2009/03/141} {\bibfield
  {journal} {\bibinfo  {journal} {JHEP}\ }\textbf {\bibinfo {volume} {03}},\
  \bibinfo {pages} {141} (\bibinfo {year} {2009})},\ \Eprint
  {http://arxiv.org/abs/0811.1024} {arXiv:0811.1024 [hep-ph]} \BibitemShut
  {NoStop}%
\bibitem [{\citenamefont {Plehn}\ \emph {et~al.}(2012)\citenamefont {Plehn},
  \citenamefont {Spannowsky},\ and\ \citenamefont {Takeuchi}}]{Plehn:2012pr}%
  \BibitemOpen
  \bibfield  {author} {\bibinfo {author} {\bibfnamefont {T.}~\bibnamefont
  {Plehn}}, \bibinfo {author} {\bibfnamefont {M.}~\bibnamefont {Spannowsky}}, \
  and\ \bibinfo {author} {\bibfnamefont {M.}~\bibnamefont {Takeuchi}},\ }\href
  {\doibase 10.1007/JHEP08(2012)091} {\bibfield  {journal} {\bibinfo  {journal}
  {JHEP}\ }\textbf {\bibinfo {volume} {08}},\ \bibinfo {pages} {091} (\bibinfo
  {year} {2012})},\ \Eprint {http://arxiv.org/abs/1205.2696} {arXiv:1205.2696
  [hep-ph]} \BibitemShut {NoStop}%
\bibitem [{\citenamefont {Han}\ \emph {et~al.}(2013)\citenamefont {Han},
  \citenamefont {Hikasa}, \citenamefont {Wu}, \citenamefont {Yang},\ and\
  \citenamefont {Zhang}}]{Han:2013kga}%
  \BibitemOpen
  \bibfield  {author} {\bibinfo {author} {\bibfnamefont {C.}~\bibnamefont
  {Han}}, \bibinfo {author} {\bibfnamefont {K.-i.}\ \bibnamefont {Hikasa}},
  \bibinfo {author} {\bibfnamefont {L.}~\bibnamefont {Wu}}, \bibinfo {author}
  {\bibfnamefont {J.~M.}\ \bibnamefont {Yang}}, \ and\ \bibinfo {author}
  {\bibfnamefont {Y.}~\bibnamefont {Zhang}},\ }\href {\doibase
  10.1007/JHEP10(2013)216} {\bibfield  {journal} {\bibinfo  {journal} {JHEP}\
  }\textbf {\bibinfo {volume} {10}},\ \bibinfo {pages} {216} (\bibinfo {year}
  {2013})},\ \Eprint {http://arxiv.org/abs/1308.5307} {arXiv:1308.5307
  [hep-ph]} \BibitemShut {NoStop}%
\bibitem [{\citenamefont {Buckley}\ \emph {et~al.}(2014)\citenamefont
  {Buckley}, \citenamefont {Plehn},\ and\ \citenamefont
  {Ramsey-Musolf}}]{Buckley:2014fqa}%
  \BibitemOpen
  \bibfield  {author} {\bibinfo {author} {\bibfnamefont {M.~R.}\ \bibnamefont
  {Buckley}}, \bibinfo {author} {\bibfnamefont {T.}~\bibnamefont {Plehn}}, \
  and\ \bibinfo {author} {\bibfnamefont {M.~J.}\ \bibnamefont
  {Ramsey-Musolf}},\ }\href {\doibase 10.1103/PhysRevD.90.014046} {\bibfield
  {journal} {\bibinfo  {journal} {Phys. Rev.}\ }\textbf {\bibinfo {volume}
  {D90}},\ \bibinfo {pages} {014046} (\bibinfo {year} {2014})},\ \Eprint
  {http://arxiv.org/abs/1403.2726} {arXiv:1403.2726 [hep-ph]} \BibitemShut
  {NoStop}%
\bibitem [{\citenamefont {Goncalves}\ \emph {et~al.}(2014)\citenamefont
  {Goncalves}, \citenamefont {Lopez-Val}, \citenamefont {Mawatari},\ and\
  \citenamefont {Plehn}}]{Goncalves:2014axa}%
  \BibitemOpen
  \bibfield  {author} {\bibinfo {author} {\bibfnamefont {D.}~\bibnamefont
  {Goncalves}}, \bibinfo {author} {\bibfnamefont {D.}~\bibnamefont
  {Lopez-Val}}, \bibinfo {author} {\bibfnamefont {K.}~\bibnamefont {Mawatari}},
  \ and\ \bibinfo {author} {\bibfnamefont {T.}~\bibnamefont {Plehn}},\ }\href
  {\doibase 10.1103/PhysRevD.90.075007} {\bibfield  {journal} {\bibinfo
  {journal} {Phys. Rev.}\ }\textbf {\bibinfo {volume} {D90}},\ \bibinfo {pages}
  {075007} (\bibinfo {year} {2014})},\ \Eprint {http://arxiv.org/abs/1407.4302}
  {arXiv:1407.4302 [hep-ph]} \BibitemShut {NoStop}%
\bibitem [{\citenamefont {Fuks}\ \emph {et~al.}(2015)\citenamefont {Fuks},
  \citenamefont {Richardson},\ and\ \citenamefont {Wilcock}}]{Fuks:2014lva}%
  \BibitemOpen
  \bibfield  {author} {\bibinfo {author} {\bibfnamefont {B.}~\bibnamefont
  {Fuks}}, \bibinfo {author} {\bibfnamefont {P.}~\bibnamefont {Richardson}}, \
  and\ \bibinfo {author} {\bibfnamefont {A.}~\bibnamefont {Wilcock}},\ }\href
  {\doibase 10.1140/epjc/s10052-015-3530-6} {\bibfield  {journal} {\bibinfo
  {journal} {Eur. Phys. J.}\ }\textbf {\bibinfo {volume} {C75}},\ \bibinfo
  {pages} {308} (\bibinfo {year} {2015})},\ \Eprint
  {http://arxiv.org/abs/1408.3634} {arXiv:1408.3634 [hep-ph]} \BibitemShut
  {NoStop}%
\bibitem [{\citenamefont {Eifert}\ and\ \citenamefont
  {Nachman}(2015)}]{Eifert:2014kea}%
  \BibitemOpen
  \bibfield  {author} {\bibinfo {author} {\bibfnamefont {T.}~\bibnamefont
  {Eifert}}\ and\ \bibinfo {author} {\bibfnamefont {B.}~\bibnamefont
  {Nachman}},\ }\href {\doibase 10.1016/j.physletb.2015.02.039} {\bibfield
  {journal} {\bibinfo  {journal} {Phys. Lett.}\ }\textbf {\bibinfo {volume}
  {B743}},\ \bibinfo {pages} {218} (\bibinfo {year} {2015})},\ \Eprint
  {http://arxiv.org/abs/1410.7025} {arXiv:1410.7025 [hep-ph]} \BibitemShut
  {NoStop}%
\bibitem [{\citenamefont {Kobakhidze}\ \emph {et~al.}(2015)\citenamefont
  {Kobakhidze}, \citenamefont {Liu}, \citenamefont {Wu},\ and\ \citenamefont
  {Yang}}]{Kobakhidze:2015dra}%
  \BibitemOpen
  \bibfield  {author} {\bibinfo {author} {\bibfnamefont {A.}~\bibnamefont
  {Kobakhidze}}, \bibinfo {author} {\bibfnamefont {N.}~\bibnamefont {Liu}},
  \bibinfo {author} {\bibfnamefont {L.}~\bibnamefont {Wu}}, \ and\ \bibinfo
  {author} {\bibfnamefont {J.~M.}\ \bibnamefont {Yang}},\ }\href {\doibase
  10.1103/PhysRevD.92.075008} {\bibfield  {journal} {\bibinfo  {journal} {Phys.
  Rev.}\ }\textbf {\bibinfo {volume} {D92}},\ \bibinfo {pages} {075008}
  (\bibinfo {year} {2015})},\ \Eprint {http://arxiv.org/abs/1504.04390}
  {arXiv:1504.04390 [hep-ph]} \BibitemShut {NoStop}%
\bibitem [{\citenamefont {Hikasa}\ \emph {et~al.}(2016)\citenamefont {Hikasa},
  \citenamefont {Li}, \citenamefont {Wu},\ and\ \citenamefont
  {Yang}}]{Hikasa:2015lma}%
  \BibitemOpen
  \bibfield  {author} {\bibinfo {author} {\bibfnamefont {K.-i.}\ \bibnamefont
  {Hikasa}}, \bibinfo {author} {\bibfnamefont {J.}~\bibnamefont {Li}}, \bibinfo
  {author} {\bibfnamefont {L.}~\bibnamefont {Wu}}, \ and\ \bibinfo {author}
  {\bibfnamefont {J.~M.}\ \bibnamefont {Yang}},\ }\href {\doibase
  10.1103/PhysRevD.93.035003} {\bibfield  {journal} {\bibinfo  {journal} {Phys.
  Rev.}\ }\textbf {\bibinfo {volume} {D93}},\ \bibinfo {pages} {035003}
  (\bibinfo {year} {2016})},\ \Eprint {http://arxiv.org/abs/1505.06006}
  {arXiv:1505.06006 [hep-ph]} \BibitemShut {NoStop}%
\bibitem [{\citenamefont {Kobakhidze}\ \emph {et~al.}(2016)\citenamefont
  {Kobakhidze}, \citenamefont {Liu}, \citenamefont {Wu}, \citenamefont {Yang},\
  and\ \citenamefont {Zhang}}]{Kobakhidze:2015scd}%
  \BibitemOpen
  \bibfield  {author} {\bibinfo {author} {\bibfnamefont {A.}~\bibnamefont
  {Kobakhidze}}, \bibinfo {author} {\bibfnamefont {N.}~\bibnamefont {Liu}},
  \bibinfo {author} {\bibfnamefont {L.}~\bibnamefont {Wu}}, \bibinfo {author}
  {\bibfnamefont {J.~M.}\ \bibnamefont {Yang}}, \ and\ \bibinfo {author}
  {\bibfnamefont {M.}~\bibnamefont {Zhang}},\ }\href {\doibase
  10.1016/j.physletb.2016.02.003} {\bibfield  {journal} {\bibinfo  {journal}
  {Phys. Lett.}\ }\textbf {\bibinfo {volume} {B755}},\ \bibinfo {pages} {76}
  (\bibinfo {year} {2016})},\ \Eprint {http://arxiv.org/abs/1511.02371}
  {arXiv:1511.02371 [hep-ph]} \BibitemShut {NoStop}%
\bibitem [{\citenamefont {Cheng}\ \emph {et~al.}(2016)\citenamefont {Cheng},
  \citenamefont {Gao}, \citenamefont {Li},\ and\ \citenamefont
  {Neill}}]{Cheng:2016mcw}%
  \BibitemOpen
  \bibfield  {author} {\bibinfo {author} {\bibfnamefont {H.-C.}\ \bibnamefont
  {Cheng}}, \bibinfo {author} {\bibfnamefont {C.}~\bibnamefont {Gao}}, \bibinfo
  {author} {\bibfnamefont {L.}~\bibnamefont {Li}}, \ and\ \bibinfo {author}
  {\bibfnamefont {N.~A.}\ \bibnamefont {Neill}},\ }\href {\doibase
  10.1007/JHEP05(2016)036} {\bibfield  {journal} {\bibinfo  {journal} {JHEP}\
  }\textbf {\bibinfo {volume} {05}},\ \bibinfo {pages} {036} (\bibinfo {year}
  {2016})},\ \Eprint {http://arxiv.org/abs/1604.00007} {arXiv:1604.00007
  [hep-ph]} \BibitemShut {NoStop}%
\bibitem [{\citenamefont {Han}\ \emph {et~al.}(2017)\citenamefont {Han},
  \citenamefont {Ren}, \citenamefont {Wu}, \citenamefont {Yang},\ and\
  \citenamefont {Zhang}}]{Han:2016xet}%
  \BibitemOpen
  \bibfield  {author} {\bibinfo {author} {\bibfnamefont {C.}~\bibnamefont
  {Han}}, \bibinfo {author} {\bibfnamefont {J.}~\bibnamefont {Ren}}, \bibinfo
  {author} {\bibfnamefont {L.}~\bibnamefont {Wu}}, \bibinfo {author}
  {\bibfnamefont {J.~M.}\ \bibnamefont {Yang}}, \ and\ \bibinfo {author}
  {\bibfnamefont {M.}~\bibnamefont {Zhang}},\ }\href {\doibase
  10.1140/epjc/s10052-017-4662-7} {\bibfield  {journal} {\bibinfo  {journal}
  {Eur. Phys. J.}\ }\textbf {\bibinfo {volume} {C77}},\ \bibinfo {pages} {93}
  (\bibinfo {year} {2017})},\ \Eprint {http://arxiv.org/abs/1609.02361}
  {arXiv:1609.02361 [hep-ph]} \BibitemShut {NoStop}%
\bibitem [{\citenamefont {Duan}\ \emph
  {et~al.}(2017{\natexlab{a}})\citenamefont {Duan}, \citenamefont {Hikasa},
  \citenamefont {Wu}, \citenamefont {Yang},\ and\ \citenamefont
  {Zhang}}]{Duan:2016vpp}%
  \BibitemOpen
  \bibfield  {author} {\bibinfo {author} {\bibfnamefont {G.~H.}\ \bibnamefont
  {Duan}}, \bibinfo {author} {\bibfnamefont {K.-i.}\ \bibnamefont {Hikasa}},
  \bibinfo {author} {\bibfnamefont {L.}~\bibnamefont {Wu}}, \bibinfo {author}
  {\bibfnamefont {J.~M.}\ \bibnamefont {Yang}}, \ and\ \bibinfo {author}
  {\bibfnamefont {M.}~\bibnamefont {Zhang}},\ }\href {\doibase
  10.1007/JHEP03(2017)091} {\bibfield  {journal} {\bibinfo  {journal} {JHEP}\
  }\textbf {\bibinfo {volume} {03}},\ \bibinfo {pages} {091} (\bibinfo {year}
  {2017}{\natexlab{a}})},\ \Eprint {http://arxiv.org/abs/1611.05211}
  {arXiv:1611.05211 [hep-ph]} \BibitemShut {NoStop}%
\bibitem [{\citenamefont {Jackson}\ \emph {et~al.}(2017)\citenamefont
  {Jackson}, \citenamefont {Rogan},\ and\ \citenamefont
  {Santoni}}]{Jackson:2016mfb}%
  \BibitemOpen
  \bibfield  {author} {\bibinfo {author} {\bibfnamefont {P.}~\bibnamefont
  {Jackson}}, \bibinfo {author} {\bibfnamefont {C.}~\bibnamefont {Rogan}}, \
  and\ \bibinfo {author} {\bibfnamefont {M.}~\bibnamefont {Santoni}},\ }\href
  {\doibase 10.1103/PhysRevD.95.035031} {\bibfield  {journal} {\bibinfo
  {journal} {Phys. Rev.}\ }\textbf {\bibinfo {volume} {D95}},\ \bibinfo {pages}
  {035031} (\bibinfo {year} {2017})},\ \Eprint
  {http://arxiv.org/abs/1607.08307} {arXiv:1607.08307 [hep-ph]} \BibitemShut
  {NoStop}%
\bibitem [{\citenamefont {Goncalves}\ \emph {et~al.}(2017)\citenamefont
  {Goncalves}, \citenamefont {Sakurai},\ and\ \citenamefont
  {Takeuchi}}]{Goncalves:2016nil}%
  \BibitemOpen
  \bibfield  {author} {\bibinfo {author} {\bibfnamefont {D.}~\bibnamefont
  {Goncalves}}, \bibinfo {author} {\bibfnamefont {K.}~\bibnamefont {Sakurai}},
  \ and\ \bibinfo {author} {\bibfnamefont {M.}~\bibnamefont {Takeuchi}},\
  }\href {\doibase 10.1103/PhysRevD.95.015030} {\bibfield  {journal} {\bibinfo
  {journal} {Phys. Rev.}\ }\textbf {\bibinfo {volume} {D95}},\ \bibinfo {pages}
  {015030} (\bibinfo {year} {2017})},\ \Eprint
  {http://arxiv.org/abs/1610.06179} {arXiv:1610.06179 [hep-ph]} \BibitemShut
  {NoStop}%
\bibitem [{\citenamefont {Butter}\ \emph {et~al.}(2018)\citenamefont {Butter},
  \citenamefont {Kasieczka}, \citenamefont {Plehn},\ and\ \citenamefont
  {Russell}}]{Butter:2017cot}%
  \BibitemOpen
  \bibfield  {author} {\bibinfo {author} {\bibfnamefont {A.}~\bibnamefont
  {Butter}}, \bibinfo {author} {\bibfnamefont {G.}~\bibnamefont {Kasieczka}},
  \bibinfo {author} {\bibfnamefont {T.}~\bibnamefont {Plehn}}, \ and\ \bibinfo
  {author} {\bibfnamefont {M.}~\bibnamefont {Russell}},\ }\href {\doibase
  10.21468/SciPostPhys.5.3.028} {\bibfield  {journal} {\bibinfo  {journal}
  {SciPost Phys.}\ }\textbf {\bibinfo {volume} {5}},\ \bibinfo {pages} {028}
  (\bibinfo {year} {2018})},\ \Eprint {http://arxiv.org/abs/1707.08966}
  {arXiv:1707.08966 [hep-ph]} \BibitemShut {NoStop}%
\bibitem [{\citenamefont {Kang}\ \emph {et~al.}(2017)\citenamefont {Kang},
  \citenamefont {Li},\ and\ \citenamefont {Zhang}}]{Kang:2017rfw}%
  \BibitemOpen
  \bibfield  {author} {\bibinfo {author} {\bibfnamefont {Z.}~\bibnamefont
  {Kang}}, \bibinfo {author} {\bibfnamefont {J.}~\bibnamefont {Li}}, \ and\
  \bibinfo {author} {\bibfnamefont {M.}~\bibnamefont {Zhang}},\ }\href
  {\doibase 10.1140/epjc/s10052-017-4951-1} {\bibfield  {journal} {\bibinfo
  {journal} {Eur. Phys. J.}\ }\textbf {\bibinfo {volume} {C77}},\ \bibinfo
  {pages} {371} (\bibinfo {year} {2017})},\ \Eprint
  {http://arxiv.org/abs/1703.08911} {arXiv:1703.08911 [hep-ph]} \BibitemShut
  {NoStop}%
\bibitem [{\citenamefont {Duan}\ \emph
  {et~al.}(2017{\natexlab{b}})\citenamefont {Duan}, \citenamefont {Wu},\ and\
  \citenamefont {Zheng}}]{Duan:2017zar}%
  \BibitemOpen
  \bibfield  {author} {\bibinfo {author} {\bibfnamefont {G.~H.}\ \bibnamefont
  {Duan}}, \bibinfo {author} {\bibfnamefont {L.}~\bibnamefont {Wu}}, \ and\
  \bibinfo {author} {\bibfnamefont {R.}~\bibnamefont {Zheng}},\ }\href
  {\doibase 10.1007/JHEP09(2017)037} {\bibfield  {journal} {\bibinfo  {journal}
  {JHEP}\ }\textbf {\bibinfo {volume} {09}},\ \bibinfo {pages} {037} (\bibinfo
  {year} {2017}{\natexlab{b}})},\ \Eprint {http://arxiv.org/abs/1706.07562}
  {arXiv:1706.07562 [hep-ph]} \BibitemShut {NoStop}%
\bibitem [{\citenamefont {Baer}\ \emph {et~al.}(2017)\citenamefont {Baer},
  \citenamefont {Barger}, \citenamefont {Gainer}, \citenamefont {Serce},\ and\
  \citenamefont {Tata}}]{Baer:2017pba}%
  \BibitemOpen
  \bibfield  {author} {\bibinfo {author} {\bibfnamefont {H.}~\bibnamefont
  {Baer}}, \bibinfo {author} {\bibfnamefont {V.}~\bibnamefont {Barger}},
  \bibinfo {author} {\bibfnamefont {J.~S.}\ \bibnamefont {Gainer}}, \bibinfo
  {author} {\bibfnamefont {H.}~\bibnamefont {Serce}}, \ and\ \bibinfo {author}
  {\bibfnamefont {X.}~\bibnamefont {Tata}},\ }\href {\doibase
  10.1103/PhysRevD.96.115008} {\bibfield  {journal} {\bibinfo  {journal} {Phys.
  Rev.}\ }\textbf {\bibinfo {volume} {D96}},\ \bibinfo {pages} {115008}
  (\bibinfo {year} {2017})},\ \Eprint {http://arxiv.org/abs/1708.09054}
  {arXiv:1708.09054 [hep-ph]} \BibitemShut {NoStop}%
\bibitem [{\citenamefont {Aaboud}\ \emph
  {et~al.}(2018{\natexlab{a}})\citenamefont {Aaboud} \emph
  {et~al.}}]{Aaboud:2018zjf}%
  \BibitemOpen
  \bibfield  {author} {\bibinfo {author} {\bibfnamefont {M.}~\bibnamefont
  {Aaboud}} \emph {et~al.} (\bibinfo {collaboration} {ATLAS}),\ }\href
  {\doibase 10.1007/JHEP09(2018)050} {\bibfield  {journal} {\bibinfo  {journal}
  {JHEP}\ }\textbf {\bibinfo {volume} {09}},\ \bibinfo {pages} {050} (\bibinfo
  {year} {2018}{\natexlab{a}})},\ \Eprint {http://arxiv.org/abs/1805.01649}
  {arXiv:1805.01649 [hep-ex]} \BibitemShut {NoStop}%
\bibitem [{\citenamefont {Aaboud}\ \emph
  {et~al.}(2017{\natexlab{a}})\citenamefont {Aaboud} \emph
  {et~al.}}]{Aaboud:2017nfd}%
  \BibitemOpen
  \bibfield  {author} {\bibinfo {author} {\bibfnamefont {M.}~\bibnamefont
  {Aaboud}} \emph {et~al.} (\bibinfo {collaboration} {ATLAS}),\ }\href
  {\doibase 10.1140/epjc/s10052-017-5445-x} {\bibfield  {journal} {\bibinfo
  {journal} {Eur. Phys. J.}\ }\textbf {\bibinfo {volume} {C77}},\ \bibinfo
  {pages} {898} (\bibinfo {year} {2017}{\natexlab{a}})},\ \Eprint
  {http://arxiv.org/abs/1708.03247} {arXiv:1708.03247 [hep-ex]} \BibitemShut
  {NoStop}%
\bibitem [{\citenamefont {Aaboud}\ \emph
  {et~al.}(2017{\natexlab{b}})\citenamefont {Aaboud} \emph
  {et~al.}}]{Aaboud:2017ayj}%
  \BibitemOpen
  \bibfield  {author} {\bibinfo {author} {\bibfnamefont {M.}~\bibnamefont
  {Aaboud}} \emph {et~al.} (\bibinfo {collaboration} {ATLAS}),\ }\href
  {\doibase 10.1007/JHEP12(2017)085} {\bibfield  {journal} {\bibinfo  {journal}
  {JHEP}\ }\textbf {\bibinfo {volume} {12}},\ \bibinfo {pages} {085} (\bibinfo
  {year} {2017}{\natexlab{b}})},\ \Eprint {http://arxiv.org/abs/1709.04183}
  {arXiv:1709.04183 [hep-ex]} \BibitemShut {NoStop}%
\bibitem [{\citenamefont {Sirunyan}\ \emph
  {et~al.}(2017{\natexlab{a}})\citenamefont {Sirunyan} \emph
  {et~al.}}]{Sirunyan:2017kqq}%
  \BibitemOpen
  \bibfield  {author} {\bibinfo {author} {\bibfnamefont {A.~M.}\ \bibnamefont
  {Sirunyan}} \emph {et~al.} (\bibinfo {collaboration} {CMS}),\ }\href
  {\doibase 10.1140/epjc/s10052-017-5267-x} {\bibfield  {journal} {\bibinfo
  {journal} {Eur. Phys. J.}\ }\textbf {\bibinfo {volume} {C77}},\ \bibinfo
  {pages} {710} (\bibinfo {year} {2017}{\natexlab{a}})},\ \Eprint
  {http://arxiv.org/abs/1705.04650} {arXiv:1705.04650 [hep-ex]} \BibitemShut
  {NoStop}%
\bibitem [{\citenamefont {Sirunyan}\ \emph
  {et~al.}(2017{\natexlab{b}})\citenamefont {Sirunyan} \emph
  {et~al.}}]{Sirunyan:2017xse}%
  \BibitemOpen
  \bibfield  {author} {\bibinfo {author} {\bibfnamefont {A.~M.}\ \bibnamefont
  {Sirunyan}} \emph {et~al.} (\bibinfo {collaboration} {CMS}),\ }\href
  {\doibase 10.1007/JHEP10(2017)019} {\bibfield  {journal} {\bibinfo  {journal}
  {JHEP}\ }\textbf {\bibinfo {volume} {10}},\ \bibinfo {pages} {019} (\bibinfo
  {year} {2017}{\natexlab{b}})},\ \Eprint {http://arxiv.org/abs/1706.04402}
  {arXiv:1706.04402 [hep-ex]} \BibitemShut {NoStop}%
\bibitem [{\citenamefont {Sirunyan}\ \emph {et~al.}(2018)\citenamefont
  {Sirunyan} \emph {et~al.}}]{Sirunyan:2018omt}%
  \BibitemOpen
  \bibfield  {author} {\bibinfo {author} {\bibfnamefont {A.~M.}\ \bibnamefont
  {Sirunyan}} \emph {et~al.} (\bibinfo {collaboration} {CMS}),\ }\href
  {\doibase 10.1007/JHEP09(2018)065} {\bibfield  {journal} {\bibinfo  {journal}
  {JHEP}\ }\textbf {\bibinfo {volume} {09}},\ \bibinfo {pages} {065} (\bibinfo
  {year} {2018})},\ \Eprint {http://arxiv.org/abs/1805.05784} {arXiv:1805.05784
  [hep-ex]} \BibitemShut {NoStop}%
\bibitem [{\citenamefont {Bhat}(2011)}]{Bhat:2010zz}%
  \BibitemOpen
  \bibfield  {author} {\bibinfo {author} {\bibfnamefont {P.~C.}\ \bibnamefont
  {Bhat}},\ }\href {\doibase 10.1146/annurev.nucl.012809.104427} {\bibfield
  {journal} {\bibinfo  {journal} {Ann. Rev. Nucl. Part. Sci.}\ }\textbf
  {\bibinfo {volume} {61}},\ \bibinfo {pages} {281} (\bibinfo {year}
  {2011})}\BibitemShut {NoStop}%
\bibitem [{\citenamefont {Roe}\ \emph {et~al.}(2005)\citenamefont {Roe},
  \citenamefont {Yang}, \citenamefont {Zhu}, \citenamefont {Liu}, \citenamefont
  {Stancu},\ and\ \citenamefont {McGregor}}]{Roe:2004na}%
  \BibitemOpen
  \bibfield  {author} {\bibinfo {author} {\bibfnamefont {B.~P.}\ \bibnamefont
  {Roe}}, \bibinfo {author} {\bibfnamefont {H.-J.}\ \bibnamefont {Yang}},
  \bibinfo {author} {\bibfnamefont {J.}~\bibnamefont {Zhu}}, \bibinfo {author}
  {\bibfnamefont {Y.}~\bibnamefont {Liu}}, \bibinfo {author} {\bibfnamefont
  {I.}~\bibnamefont {Stancu}}, \ and\ \bibinfo {author} {\bibfnamefont
  {G.}~\bibnamefont {McGregor}},\ }\href {\doibase 10.1016/j.nima.2004.12.018}
  {\bibfield  {journal} {\bibinfo  {journal} {Nucl. Instrum. Meth.}\ }\textbf
  {\bibinfo {volume} {A543}},\ \bibinfo {pages} {577} (\bibinfo {year}
  {2005})},\ \Eprint {http://arxiv.org/abs/physics/0408124}
  {arXiv:physics/0408124 [physics]} \BibitemShut {NoStop}%
\bibitem [{\citenamefont {Baldi}\ \emph {et~al.}(2014)\citenamefont {Baldi},
  \citenamefont {Sadowski},\ and\ \citenamefont {Whiteson}}]{Baldi:2014kfa}%
  \BibitemOpen
  \bibfield  {author} {\bibinfo {author} {\bibfnamefont {P.}~\bibnamefont
  {Baldi}}, \bibinfo {author} {\bibfnamefont {P.}~\bibnamefont {Sadowski}}, \
  and\ \bibinfo {author} {\bibfnamefont {D.}~\bibnamefont {Whiteson}},\ }\href
  {\doibase 10.1038/ncomms5308} {\bibfield  {journal} {\bibinfo  {journal}
  {Nature Commun.}\ }\textbf {\bibinfo {volume} {5}},\ \bibinfo {pages} {4308}
  (\bibinfo {year} {2014})},\ \Eprint {http://arxiv.org/abs/1402.4735}
  {arXiv:1402.4735 [hep-ph]} \BibitemShut {NoStop}%
\bibitem [{\citenamefont {Baldi}\ \emph {et~al.}(2015)\citenamefont {Baldi},
  \citenamefont {Sadowski},\ and\ \citenamefont {Whiteson}}]{Baldi:2014pta}%
  \BibitemOpen
  \bibfield  {author} {\bibinfo {author} {\bibfnamefont {P.}~\bibnamefont
  {Baldi}}, \bibinfo {author} {\bibfnamefont {P.}~\bibnamefont {Sadowski}}, \
  and\ \bibinfo {author} {\bibfnamefont {D.}~\bibnamefont {Whiteson}},\ }\href
  {\doibase 10.1103/PhysRevLett.114.111801} {\bibfield  {journal} {\bibinfo
  {journal} {Phys. Rev. Lett.}\ }\textbf {\bibinfo {volume} {114}},\ \bibinfo
  {pages} {111801} (\bibinfo {year} {2015})},\ \Eprint
  {http://arxiv.org/abs/1410.3469} {arXiv:1410.3469 [hep-ph]} \BibitemShut
  {NoStop}%
\bibitem [{\citenamefont {Bridges}\ \emph {et~al.}(2011)\citenamefont
  {Bridges}, \citenamefont {Cranmer}, \citenamefont {Feroz}, \citenamefont
  {Hobson}, \citenamefont {Ruiz~de Austri},\ and\ \citenamefont
  {Trotta}}]{Bridges:2010de}%
  \BibitemOpen
  \bibfield  {author} {\bibinfo {author} {\bibfnamefont {M.}~\bibnamefont
  {Bridges}}, \bibinfo {author} {\bibfnamefont {K.}~\bibnamefont {Cranmer}},
  \bibinfo {author} {\bibfnamefont {F.}~\bibnamefont {Feroz}}, \bibinfo
  {author} {\bibfnamefont {M.}~\bibnamefont {Hobson}}, \bibinfo {author}
  {\bibfnamefont {R.}~\bibnamefont {Ruiz~de Austri}}, \ and\ \bibinfo {author}
  {\bibfnamefont {R.}~\bibnamefont {Trotta}},\ }\href {\doibase
  10.1007/JHEP03(2011)012} {\bibfield  {journal} {\bibinfo  {journal} {JHEP}\
  }\textbf {\bibinfo {volume} {03}},\ \bibinfo {pages} {012} (\bibinfo {year}
  {2011})},\ \Eprint {http://arxiv.org/abs/1011.4306} {arXiv:1011.4306
  [hep-ph]} \BibitemShut {NoStop}%
\bibitem [{\citenamefont {Buckley}\ \emph {et~al.}(2012)\citenamefont
  {Buckley}, \citenamefont {Shilton},\ and\ \citenamefont
  {White}}]{Buckley:2011kc}%
  \BibitemOpen
  \bibfield  {author} {\bibinfo {author} {\bibfnamefont {A.}~\bibnamefont
  {Buckley}}, \bibinfo {author} {\bibfnamefont {A.}~\bibnamefont {Shilton}}, \
  and\ \bibinfo {author} {\bibfnamefont {M.~J.}\ \bibnamefont {White}},\ }\href
  {\doibase 10.1016/j.cpc.2011.12.026} {\bibfield  {journal} {\bibinfo
  {journal} {Comput. Phys. Commun.}\ }\textbf {\bibinfo {volume} {183}},\
  \bibinfo {pages} {960} (\bibinfo {year} {2012})},\ \Eprint
  {http://arxiv.org/abs/1106.4613} {arXiv:1106.4613 [hep-ph]} \BibitemShut
  {NoStop}%
\bibitem [{\citenamefont {Bornhauser}\ and\ \citenamefont
  {Drees}(2013)}]{Bornhauser:2013aya}%
  \BibitemOpen
  \bibfield  {author} {\bibinfo {author} {\bibfnamefont {N.}~\bibnamefont
  {Bornhauser}}\ and\ \bibinfo {author} {\bibfnamefont {M.}~\bibnamefont
  {Drees}},\ }\href {\doibase 10.1103/PhysRevD.88.075016} {\bibfield  {journal}
  {\bibinfo  {journal} {Phys. Rev.}\ }\textbf {\bibinfo {volume} {D88}},\
  \bibinfo {pages} {075016} (\bibinfo {year} {2013})},\ \Eprint
  {http://arxiv.org/abs/1307.3383} {arXiv:1307.3383 [hep-ph]} \BibitemShut
  {NoStop}%
\bibitem [{\citenamefont {Caron}\ \emph {et~al.}(2017)\citenamefont {Caron},
  \citenamefont {Kim}, \citenamefont {Rolbiecki}, \citenamefont {Ruiz~de
  Austri},\ and\ \citenamefont {Stienen}}]{Caron:2016hib}%
  \BibitemOpen
  \bibfield  {author} {\bibinfo {author} {\bibfnamefont {S.}~\bibnamefont
  {Caron}}, \bibinfo {author} {\bibfnamefont {J.~S.}\ \bibnamefont {Kim}},
  \bibinfo {author} {\bibfnamefont {K.}~\bibnamefont {Rolbiecki}}, \bibinfo
  {author} {\bibfnamefont {R.}~\bibnamefont {Ruiz~de Austri}}, \ and\ \bibinfo
  {author} {\bibfnamefont {B.}~\bibnamefont {Stienen}},\ }\href {\doibase
  10.1140/epjc/s10052-017-4814-9} {\bibfield  {journal} {\bibinfo  {journal}
  {Eur. Phys. J.}\ }\textbf {\bibinfo {volume} {C77}},\ \bibinfo {pages} {257}
  (\bibinfo {year} {2017})},\ \Eprint {http://arxiv.org/abs/1605.02797}
  {arXiv:1605.02797 [hep-ph]} \BibitemShut {NoStop}%
\bibitem [{\citenamefont {Bertone}\ \emph {et~al.}(2016)\citenamefont
  {Bertone}, \citenamefont {Deisenroth}, \citenamefont {Kim}, \citenamefont
  {Liem}, \citenamefont {Ruiz~de Austri},\ and\ \citenamefont
  {Welling}}]{Bertone:2016mdy}%
  \BibitemOpen
  \bibfield  {author} {\bibinfo {author} {\bibfnamefont {G.}~\bibnamefont
  {Bertone}}, \bibinfo {author} {\bibfnamefont {M.~P.}\ \bibnamefont
  {Deisenroth}}, \bibinfo {author} {\bibfnamefont {J.~S.}\ \bibnamefont {Kim}},
  \bibinfo {author} {\bibfnamefont {S.}~\bibnamefont {Liem}}, \bibinfo {author}
  {\bibfnamefont {R.}~\bibnamefont {Ruiz~de Austri}}, \ and\ \bibinfo {author}
  {\bibfnamefont {M.}~\bibnamefont {Welling}},\ }\href@noop {} {\  (\bibinfo
  {year} {2016})},\ \Eprint {http://arxiv.org/abs/1611.02704} {arXiv:1611.02704
  [hep-ph]} \BibitemShut {NoStop}%
\bibitem [{\citenamefont {{Gilmer}}\ \emph {et~al.}(2017)\citenamefont
  {{Gilmer}}, \citenamefont {{Schoenholz}}, \citenamefont {{Riley}},
  \citenamefont {{Vinyals}},\ and\ \citenamefont
  {{Dahl}}}]{2017arXiv170401212G}%
  \BibitemOpen
  \bibfield  {author} {\bibinfo {author} {\bibfnamefont {J.}~\bibnamefont
  {{Gilmer}}}, \bibinfo {author} {\bibfnamefont {S.~S.}\ \bibnamefont
  {{Schoenholz}}}, \bibinfo {author} {\bibfnamefont {P.~F.}\ \bibnamefont
  {{Riley}}}, \bibinfo {author} {\bibfnamefont {O.}~\bibnamefont {{Vinyals}}},
  \ and\ \bibinfo {author} {\bibfnamefont {G.~E.}\ \bibnamefont {{Dahl}}},\
  }\href@noop {} {\bibfield  {journal} {\bibinfo  {journal} {ArXiv e-prints}\ }
  (\bibinfo {year} {2017})},\ \Eprint {http://arxiv.org/abs/1704.01212}
  {arXiv:1704.01212 [cs.LG]} \BibitemShut {NoStop}%
\bibitem [{\citenamefont {Gori}\ \emph {et~al.}(2005)\citenamefont {Gori},
  \citenamefont {Monfardini},\ and\ \citenamefont {Scarselli}}]{Gori05}%
  \BibitemOpen
  \bibfield  {author} {\bibinfo {author} {\bibfnamefont {M.}~\bibnamefont
  {Gori}}, \bibinfo {author} {\bibfnamefont {G.}~\bibnamefont {Monfardini}}, \
  and\ \bibinfo {author} {\bibfnamefont {F.}~\bibnamefont {Scarselli}},\ }in\
  \href {\doibase 10.1109/IJCNN.2005.1555942} {\emph {\bibinfo {booktitle}
  {Proceedings. 2005 IEEE International Joint Conference on Neural Networks,
  2005.}}},\ Vol.~\bibinfo {volume} {2}\ (\bibinfo {year} {2005})\ pp.\
  \bibinfo {pages} {729--734 vol. 2}\BibitemShut {NoStop}%
\bibitem [{\citenamefont {Scarselli}\ \emph {et~al.}(2009)\citenamefont
  {Scarselli}, \citenamefont {Gori}, \citenamefont {Tsoi}, \citenamefont
  {Hagenbuchner},\ and\ \citenamefont {Monfardini}}]{Scarselli09}%
  \BibitemOpen
  \bibfield  {author} {\bibinfo {author} {\bibfnamefont {F.}~\bibnamefont
  {Scarselli}}, \bibinfo {author} {\bibfnamefont {M.}~\bibnamefont {Gori}},
  \bibinfo {author} {\bibfnamefont {A.~C.}\ \bibnamefont {Tsoi}}, \bibinfo
  {author} {\bibfnamefont {M.}~\bibnamefont {Hagenbuchner}}, \ and\ \bibinfo
  {author} {\bibfnamefont {G.}~\bibnamefont {Monfardini}},\ }\href {\doibase
  10.1109/TNN.2008.2005605} {\bibfield  {journal} {\bibinfo  {journal} {IEEE
  Transactions on Neural Networks}\ }\textbf {\bibinfo {volume} {20}},\
  \bibinfo {pages} {61} (\bibinfo {year} {2009})}\BibitemShut {NoStop}%
\bibitem [{\citenamefont {Henrion}\ \emph {et~al.}(2017)\citenamefont
  {Henrion}, \citenamefont {Cranmer}, \citenamefont {Bruna}, \citenamefont
  {Cho}, \citenamefont {Brehmer}, \citenamefont {Louppe},\ and\ \citenamefont
  {Rochette}}]{Henrion17}%
  \BibitemOpen
  \bibfield  {author} {\bibinfo {author} {\bibfnamefont {I.}~\bibnamefont
  {Henrion}}, \bibinfo {author} {\bibfnamefont {K.}~\bibnamefont {Cranmer}},
  \bibinfo {author} {\bibfnamefont {J.}~\bibnamefont {Bruna}}, \bibinfo
  {author} {\bibfnamefont {K.}~\bibnamefont {Cho}}, \bibinfo {author}
  {\bibfnamefont {J.}~\bibnamefont {Brehmer}}, \bibinfo {author} {\bibfnamefont
  {G.}~\bibnamefont {Louppe}}, \ and\ \bibinfo {author} {\bibfnamefont
  {G.}~\bibnamefont {Rochette}},\ }in\ \href@noop {} {\emph {\bibinfo
  {booktitle} {Deep Learning for Physical Sciences. Workshop at the 31st
  Conference on Neural Information Processing Systems (NIPS), 2017.}}}\
  (\bibinfo {year} {2017})\BibitemShut {NoStop}%
\bibitem [{\citenamefont {Kingma}\ and\ \citenamefont
  {Ba}(2014)}]{DBLP:journals/corr/KingmaB14}%
  \BibitemOpen
  \bibfield  {author} {\bibinfo {author} {\bibfnamefont {D.~P.}\ \bibnamefont
  {Kingma}}\ and\ \bibinfo {author} {\bibfnamefont {J.}~\bibnamefont {Ba}},\
  }\href {http://arxiv.org/abs/1412.6980} {\bibfield  {journal} {\bibinfo
  {journal} {CoRR}\ }\textbf {\bibinfo {volume} {abs/1412.6980}} (\bibinfo
  {year} {2014})},\ \Eprint {http://arxiv.org/abs/1412.6980} {arXiv:1412.6980}
  \BibitemShut {NoStop}%
\bibitem [{\citenamefont {http://pytorch.org/}()}]{pytorch}%
  \BibitemOpen
  \bibfield  {author} {\bibinfo {author} {\bibnamefont {http://pytorch.org/}},\
  }\href {http://pytorch.org/} {}\BibitemShut {NoStop}%
\bibitem [{\citenamefont {Aaboud}\ \emph
  {et~al.}(2018{\natexlab{b}})\citenamefont {Aaboud} \emph
  {et~al.}}]{Aaboud:2017aeu}%
  \BibitemOpen
  \bibfield  {author} {\bibinfo {author} {\bibfnamefont {M.}~\bibnamefont
  {Aaboud}} \emph {et~al.} (\bibinfo {collaboration} {ATLAS}),\ }\href
  {\doibase 10.1007/JHEP06(2018)108} {\bibfield  {journal} {\bibinfo  {journal}
  {JHEP}\ }\textbf {\bibinfo {volume} {06}},\ \bibinfo {pages} {108} (\bibinfo
  {year} {2018}{\natexlab{b}})},\ \Eprint {http://arxiv.org/abs/1711.11520}
  {arXiv:1711.11520 [hep-ex]} \BibitemShut {NoStop}%
\bibitem [{\citenamefont {Alwall}\ \emph {et~al.}(2014)\citenamefont {Alwall},
  \citenamefont {Frederix}, \citenamefont {Frixione}, \citenamefont {Hirschi},
  \citenamefont {Maltoni}, \citenamefont {Mattelaer}, \citenamefont {Shao},
  \citenamefont {Stelzer}, \citenamefont {Torrielli},\ and\ \citenamefont
  {Zaro}}]{Alwall:2014hca}%
  \BibitemOpen
  \bibfield  {author} {\bibinfo {author} {\bibfnamefont {J.}~\bibnamefont
  {Alwall}}, \bibinfo {author} {\bibfnamefont {R.}~\bibnamefont {Frederix}},
  \bibinfo {author} {\bibfnamefont {S.}~\bibnamefont {Frixione}}, \bibinfo
  {author} {\bibfnamefont {V.}~\bibnamefont {Hirschi}}, \bibinfo {author}
  {\bibfnamefont {F.}~\bibnamefont {Maltoni}}, \bibinfo {author} {\bibfnamefont
  {O.}~\bibnamefont {Mattelaer}}, \bibinfo {author} {\bibfnamefont {H.~S.}\
  \bibnamefont {Shao}}, \bibinfo {author} {\bibfnamefont {T.}~\bibnamefont
  {Stelzer}}, \bibinfo {author} {\bibfnamefont {P.}~\bibnamefont {Torrielli}},
  \ and\ \bibinfo {author} {\bibfnamefont {M.}~\bibnamefont {Zaro}},\ }\href
  {\doibase 10.1007/JHEP07(2014)079} {\bibfield  {journal} {\bibinfo  {journal}
  {JHEP}\ }\textbf {\bibinfo {volume} {07}},\ \bibinfo {pages} {079} (\bibinfo
  {year} {2014})},\ \Eprint {http://arxiv.org/abs/1405.0301} {arXiv:1405.0301
  [hep-ph]} \BibitemShut {NoStop}%
\bibitem [{\citenamefont {Sjöstrand}\ \emph {et~al.}(2015)\citenamefont
  {Sjöstrand}, \citenamefont {Ask}, \citenamefont {Christiansen},
  \citenamefont {Corke}, \citenamefont {Desai}, \citenamefont {Ilten},
  \citenamefont {Mrenna}, \citenamefont {Prestel}, \citenamefont {Rasmussen},\
  and\ \citenamefont {Skands}}]{Sjostrand:2014zea}%
  \BibitemOpen
  \bibfield  {author} {\bibinfo {author} {\bibfnamefont {T.}~\bibnamefont
  {Sjöstrand}}, \bibinfo {author} {\bibfnamefont {S.}~\bibnamefont {Ask}},
  \bibinfo {author} {\bibfnamefont {J.~R.}\ \bibnamefont {Christiansen}},
  \bibinfo {author} {\bibfnamefont {R.}~\bibnamefont {Corke}}, \bibinfo
  {author} {\bibfnamefont {N.}~\bibnamefont {Desai}}, \bibinfo {author}
  {\bibfnamefont {P.}~\bibnamefont {Ilten}}, \bibinfo {author} {\bibfnamefont
  {S.}~\bibnamefont {Mrenna}}, \bibinfo {author} {\bibfnamefont
  {S.}~\bibnamefont {Prestel}}, \bibinfo {author} {\bibfnamefont {C.~O.}\
  \bibnamefont {Rasmussen}}, \ and\ \bibinfo {author} {\bibfnamefont {P.~Z.}\
  \bibnamefont {Skands}},\ }\href {\doibase 10.1016/j.cpc.2015.01.024}
  {\bibfield  {journal} {\bibinfo  {journal} {Comput. Phys. Commun.}\ }\textbf
  {\bibinfo {volume} {191}},\ \bibinfo {pages} {159} (\bibinfo {year}
  {2015})},\ \Eprint {http://arxiv.org/abs/1410.3012} {arXiv:1410.3012
  [hep-ph]} \BibitemShut {NoStop}%
\bibitem [{\citenamefont {de~Favereau}\ \emph {et~al.}(2014)\citenamefont
  {de~Favereau}, \citenamefont {Delaere}, \citenamefont {Demin}, \citenamefont
  {Giammanco}, \citenamefont {Lemaître}, \citenamefont {Mertens},\ and\
  \citenamefont {Selvaggi}}]{deFavereau:2013fsa}%
  \BibitemOpen
  \bibfield  {author} {\bibinfo {author} {\bibfnamefont {J.}~\bibnamefont
  {de~Favereau}}, \bibinfo {author} {\bibfnamefont {C.}~\bibnamefont
  {Delaere}}, \bibinfo {author} {\bibfnamefont {P.}~\bibnamefont {Demin}},
  \bibinfo {author} {\bibfnamefont {A.}~\bibnamefont {Giammanco}}, \bibinfo
  {author} {\bibfnamefont {V.}~\bibnamefont {Lemaître}}, \bibinfo {author}
  {\bibfnamefont {A.}~\bibnamefont {Mertens}}, \ and\ \bibinfo {author}
  {\bibfnamefont {M.}~\bibnamefont {Selvaggi}} (\bibinfo {collaboration}
  {DELPHES 3}),\ }\href {\doibase 10.1007/JHEP02(2014)057} {\bibfield
  {journal} {\bibinfo  {journal} {JHEP}\ }\textbf {\bibinfo {volume} {02}},\
  \bibinfo {pages} {057} (\bibinfo {year} {2014})},\ \Eprint
  {http://arxiv.org/abs/1307.6346} {arXiv:1307.6346 [hep-ex]} \BibitemShut
  {NoStop}%
\bibitem [{\citenamefont {Cacciari}\ \emph {et~al.}(2008)\citenamefont
  {Cacciari}, \citenamefont {Salam},\ and\ \citenamefont
  {Soyez}}]{Cacciari:2008gp}%
  \BibitemOpen
  \bibfield  {author} {\bibinfo {author} {\bibfnamefont {M.}~\bibnamefont
  {Cacciari}}, \bibinfo {author} {\bibfnamefont {G.~P.}\ \bibnamefont {Salam}},
  \ and\ \bibinfo {author} {\bibfnamefont {G.}~\bibnamefont {Soyez}},\ }\href
  {\doibase 10.1088/1126-6708/2008/04/063} {\bibfield  {journal} {\bibinfo
  {journal} {JHEP}\ }\textbf {\bibinfo {volume} {04}},\ \bibinfo {pages} {063}
  (\bibinfo {year} {2008})},\ \Eprint {http://arxiv.org/abs/0802.1189}
  {arXiv:0802.1189 [hep-ph]} \BibitemShut {NoStop}%
\bibitem [{\citenamefont {Drees}\ \emph {et~al.}(2015)\citenamefont {Drees},
  \citenamefont {Dreiner}, \citenamefont {Schmeier}, \citenamefont
  {Tattersall},\ and\ \citenamefont {Kim}}]{Drees:2013wra}%
  \BibitemOpen
  \bibfield  {author} {\bibinfo {author} {\bibfnamefont {M.}~\bibnamefont
  {Drees}}, \bibinfo {author} {\bibfnamefont {H.}~\bibnamefont {Dreiner}},
  \bibinfo {author} {\bibfnamefont {D.}~\bibnamefont {Schmeier}}, \bibinfo
  {author} {\bibfnamefont {J.}~\bibnamefont {Tattersall}}, \ and\ \bibinfo
  {author} {\bibfnamefont {J.~S.}\ \bibnamefont {Kim}},\ }\href {\doibase
  10.1016/j.cpc.2014.10.018} {\bibfield  {journal} {\bibinfo  {journal}
  {Comput. Phys. Commun.}\ }\textbf {\bibinfo {volume} {187}},\ \bibinfo
  {pages} {227} (\bibinfo {year} {2015})},\ \Eprint
  {http://arxiv.org/abs/1312.2591} {arXiv:1312.2591 [hep-ph]} \BibitemShut
  {NoStop}%
\bibitem [{\citenamefont {Beenakker}\ \emph {et~al.}(1999)\citenamefont
  {Beenakker}, \citenamefont {Klasen}, \citenamefont {Kramer}, \citenamefont
  {Plehn}, \citenamefont {Spira},\ and\ \citenamefont
  {Zerwas}}]{Beenakker:1999xh}%
  \BibitemOpen
  \bibfield  {author} {\bibinfo {author} {\bibfnamefont {W.}~\bibnamefont
  {Beenakker}}, \bibinfo {author} {\bibfnamefont {M.}~\bibnamefont {Klasen}},
  \bibinfo {author} {\bibfnamefont {M.}~\bibnamefont {Kramer}}, \bibinfo
  {author} {\bibfnamefont {T.}~\bibnamefont {Plehn}}, \bibinfo {author}
  {\bibfnamefont {M.}~\bibnamefont {Spira}}, \ and\ \bibinfo {author}
  {\bibfnamefont {P.~M.}\ \bibnamefont {Zerwas}},\ }\href {\doibase
  10.1103/PhysRevLett.100.029901, 10.1103/PhysRevLett.83.3780} {\bibfield
  {journal} {\bibinfo  {journal} {Phys. Rev. Lett.}\ }\textbf {\bibinfo
  {volume} {83}},\ \bibinfo {pages} {3780} (\bibinfo {year} {1999})},\ \bibinfo
  {note} {[Erratum: Phys. Rev. Lett.100,029901(2008)]},\ \Eprint
  {http://arxiv.org/abs/hep-ph/9906298} {arXiv:hep-ph/9906298 [hep-ph]}
  \BibitemShut {NoStop}%
\bibitem [{\citenamefont {Czakon}\ and\ \citenamefont
  {Mitov}(2014)}]{Czakon:2011xx}%
  \BibitemOpen
  \bibfield  {author} {\bibinfo {author} {\bibfnamefont {M.}~\bibnamefont
  {Czakon}}\ and\ \bibinfo {author} {\bibfnamefont {A.}~\bibnamefont {Mitov}},\
  }\href {\doibase 10.1016/j.cpc.2014.06.021} {\bibfield  {journal} {\bibinfo
  {journal} {Comput. Phys. Commun.}\ }\textbf {\bibinfo {volume} {185}},\
  \bibinfo {pages} {2930} (\bibinfo {year} {2014})},\ \Eprint
  {http://arxiv.org/abs/1112.5675} {arXiv:1112.5675 [hep-ph]} \BibitemShut
  {NoStop}%
\bibitem [{\citenamefont {Boughezal}\ \emph {et~al.}(2015)\citenamefont
  {Boughezal}, \citenamefont {Focke}, \citenamefont {Liu},\ and\ \citenamefont
  {Petriello}}]{Boughezal:2015dva}%
  \BibitemOpen
  \bibfield  {author} {\bibinfo {author} {\bibfnamefont {R.}~\bibnamefont
  {Boughezal}}, \bibinfo {author} {\bibfnamefont {C.}~\bibnamefont {Focke}},
  \bibinfo {author} {\bibfnamefont {X.}~\bibnamefont {Liu}}, \ and\ \bibinfo
  {author} {\bibfnamefont {F.}~\bibnamefont {Petriello}},\ }\href {\doibase
  10.1103/PhysRevLett.115.062002} {\bibfield  {journal} {\bibinfo  {journal}
  {Phys. Rev. Lett.}\ }\textbf {\bibinfo {volume} {115}},\ \bibinfo {pages}
  {062002} (\bibinfo {year} {2015})},\ \Eprint
  {http://arxiv.org/abs/1504.02131} {arXiv:1504.02131 [hep-ph]} \BibitemShut
  {NoStop}%
\end{thebibliography}

%

\end{document}